\documentclass[12pt]{article}
\usepackage[margin=1in]{geometry}
\usepackage{authblk}
\usepackage{babel}
\usepackage{mathtools}
\usepackage{amssymb} 
\usepackage{float}
\usepackage{enumitem}
\usepackage{amsthm}
\usepackage{amsmath}
\usepackage{amsfonts}
\usepackage{array}
\usepackage{amsopn}
\usepackage{multirow}
\usepackage{longtable}
 \usepackage[table,xcdraw]{xcolor}
\usepackage[numbers]{natbib}
\usepackage{enumitem}
\DeclarePairedDelimiterX\Basics[1](){ #1}
\usepackage[utf8]{inputenc}
\usepackage{color}
\usepackage{caption}
\usepackage{subfigure}
\usepackage{enumitem}
\usepackage{pifont}
\usepackage{algorithm, algpseudocode, algcompatible}
\usepackage[T1]{fontenc}
\usepackage{etoolbox}
\usepackage{arydshln}
\usepackage{lscape}
\usepackage{bm}
\usepackage{multirow}
\usepackage{lineno}
\usepackage{diagbox}
\usepackage[flushleft]{threeparttable}
\usepackage{array, booktabs, multirow}
\usepackage{graphicx}
\usepackage{xcolor}

\definecolor{colorA}{rgb}{0.122, 0.467, 0.706}
\definecolor{colorB}{rgb}{1.0, 0.498, 0.055}
\definecolor{colorC}{rgb}{0.173, 0.627, 0.173}
\definecolor{colorD}{rgb}{0.839, 0.153, 0.157}
\definecolor{colorE}{rgb}{0.58, 0.404, 0.741}
\definecolor{colorF}{rgb}{0.549, 0.337, 0.294}
\definecolor{colorG}{rgb}{0.89, 0.467, 0.761}
\definecolor{colorH}{rgb}{0.498, 0.498, 0.498}
\definecolor{colorI}{rgb}{0.737, 0.741, 0.133}
\definecolor{colorJ}{rgb}{0.09, 0.745, 0.812}

\newcounter{noteRCctr} 

\newcounter{noteMCctr} 

\definecolor{colour3}{RGB}{178,55,250} 

\newcounter{noteCZctr} 

\usepackage[colorlinks=true,linkcolor=red,filecolor=green,citecolor=blue]{hyperref}

\title{\textbf{Cross-Impact of Order Flow Imbalance \\ in Equity Markets
}}

\author[1,5]{Rama Cont}
\author[1,2,3,4]{Mihai Cucuringu}
\author[1,2,3]{Chao Zhang\thanks{Corresponding author.
Email: chao.zhang@stats.ox.co.uk. We thank Robert Engle, {\'A}lvaro Cartea, Slavi Marinov and seminar participants at the 11th Bachelier World Congress 2022; the 14th Annual SoFiE Conference; University of Oxford for helpful comments. We also thank the Oxford Suzhou Centre for Advanced Research for providing the computational facilities. Earlier versions of this paper circulated under the titles ``Price Impact of Order Flow Imbalance: Multi-level, Cross-asset and Forecasting'' and ``Cross-Impact of Order Flow Imbalance: Contemporaneous and Predictive''. First draft: December 2021.}}

\affil[1]{\small Mathematical Institute, University of Oxford, Oxford, UK}
\affil[2]{\small Department of Statistics, University of Oxford, Oxford, UK}
\affil[3]{\small Oxford-Man Institute of Quantitative Finance, University of Oxford, Oxford, UK}
\affil[4]{\small The Alan Turing Institute, London, UK}
\affil[5]{\small Oxford Suzhou Centre for Advanced Research, Suzhou, China}
\date{May 2023}

\begin{document}

\maketitle
\begin{abstract}
\noindent

We investigate the impact of order flow imbalance (OFI) on price movements in equity markets in a multi-asset setting. First, we propose a systematic approach for combining OFIs at the top levels of the limit order book into an integrated OFI variable which better explains price impact, compared to the best-level OFI. We show that  once the information from multiple levels is integrated into   OFI, multi-asset models with cross-impact do not provide additional explanatory power for contemporaneous impact compared to a sparse model without   cross-impact terms.  On the other hand, we show that lagged cross-asset OFIs do improve the forecasting of future returns. We also establish that this lagged cross-impact mainly manifests at short-term horizons and decays rapidly in time.

\end{abstract}


\newpage
\tableofcontents
\newpage

\newpage

\section{Introduction}

Accurately estimating and forecasting the impact of trading behavior of market participants on the price movements of assets carries practical implications for both practitioners and academics, such as trading cost analysis (\citet{frazzini2012trading}) and optimal execution of trades. The impact of  trades on asset  prices,  known as {\it price impact}, has been the focus of many studies and modeling efforts (\citet{cont2014price, lillo2003master}). In a multi-asset setting, several studied have focused on the concept of \textbf{cross-impact}, which attempts to describe the impact of trading a given asset on the price of \textit{other} assets  (see \citet{benzaquen2017dissecting, capponi2020multi, pasquariello2015strategic}). 

{Several studies have investigated  \textit{contemporaneous cross-impact} of order flow on returns   by examining their  cross-correlation structure.} For example, \citet{hasbrouck2001common} revealed that commonality in returns among Dow 30 stocks is mostly attributed {to} order flow commonality. \citet{tomas2022build} built a principled approach to choosing a cross-impact model for various markets.  \citet{capponi2020multi} showed that the positive covariance between returns of a specific stock and order flow imbalances of other stocks does not necessarily constitute evidence of cross-impact. They further demonstrated that, as long as the common factor in order flow imbalances is taken into account, adding cross-impact terms only marginally improves model performance, {and thus  may be disregarded}.  
Our  study complements \citet{capponi2020multi} in several ways:
unlike \cite{capponi2020multi} which focuses on  in-sample performance, we also consider the forecasting power of cross-order flow using both single and multi-level OFIs. To the best of our knowledge, there have been no studies that examine the influence of order flows on price movements in a multi-asset setting, while also taking into account the deeper levels in the limit order book (LOB).\footnote{\citet{xu2018multi} studied the contemporaneous price impact (not cross-impact) model by extending the model of \citet{cont2014price} to multi-level order flow imbalance.}

A more challenging problem than explaining contemporaneous returns is to examine the impact of trade orders on prices over future horizons, which has received a lot less attention in the literature, {despite its important economic implications}. Some studies have examined the relationship between order imbalances and \textit{future daily} returns.\footnote{Several studies, such as \citet{hou2007industry, menzly2010market, chinco2019sparse, buccheri2021high}, investigated the lead-lag effect in equity returns across various assets, but did not take into account order flows.} \citet{chordia2002order} revealed that daily stock market returns are strongly related to contemporaneous and lagged order imbalances. \citet{chordia2004order} further found that there exists a positive relation between lagged order imbalances and daily individual stock returns. The authors also showed that imbalance-based trading strategies, i.e. buy if the previous day's imbalance is positive, and sell if the previous day's imbalance is negative, are able to yield statistically significant profits.
\citet{pasquariello2015strategic} provided empirical evidence of cross-asset informational effects in NYSE and NASDAQ stocks between 1993 and 2004, and demonstrated that the daily order flow imbalance in one stock, or across one industry, has a significant and persistent impact on daily returns of other stocks or industries. {\citet{rosenbaum2021characterisation} provided a characterization of the class of cross-impact kernels for a market that employs Hawkes processes to model trades and applied their method to two instruments from E-Mini Futures.}

{Given the recent progress in high-frequency trading (HFT), it is increasingly crucial to obtain accurate estimations of the cross-impact on \textit{future intraday} returns.
\citet{benzaquen2017dissecting} introduced a multivariate linear  model (see \citet{kyle1985continuous}) to describe the structure of cross-impact and found that a significant fraction of the covariance of stock returns can be accounted for by this model. \citet{wang2016average, wang2018statistical} empirically analyzed and discussed the impact of trading a specific stock on the average price change of the whole market or {of individual} sectors. 
\citet{schneider2019cross} derived theoretical limits for the size and form of cross-impact and verified them on sovereign bonds data. However, when modeling cross-impact, these methods do not consider the possibility of high correlations between cross-asset order flows, which may result in overfitting issues. This is also evidenced by studies such as  \citet{benzaquen2017dissecting} and \citet{tomas2022build}. Moreover, these studies mainly investigated the cross-impact coefficients for a fixed time period (i.e., in a static setting), ignoring the temporal dynamics of cross-impact.}

In recent years, machine learning models including deep neural networks, have achieved substantial developments, leading to their applications in financial markets, especially for the task of modeling stock returns.  For example, \citet{huck2019large} utilized state-of-the-art techniques, such as random forests, to construct a portfolio over a period of 22 years, and the results demonstrated the power of machine learning models to produce profitable trading signals. \citet{krauss2017deep} applied a series of machine learning methods to forecast the probability of a stock outperforming the market index, and then constructed long-short portfolios from the predicted one-day-ahead trading signals. \citet{gu2020empirical} employed a set of machine learning methods to make one-month-ahead return forecasts, and demonstrated the potential of machine learning approaches in empirical asset pricing, due to their ability to handle nonlinear interactions. \citet{ait2022and} investigated the predictability of high-frequency stock returns and durations using LASSO and tree methods via many relevant predictors derived from returns and order flows. \citet{tashiro2019encoding} and \citet{kolm2021deep} applied deep neural networks with LOB-based features to predict high-frequency returns. Nonetheless, to the best of our knowledge, cross-asset order flow imbalances have not been considered as predictors for forecasting future high-frequency returns in the literature, which is one of the main directions we explore in the second half of this paper.

\subsection{Main contributions}
The present study makes two main contributions to the literature regarding the \textit{contemporaneous} and \textit{predictive} cross-impact of order flow imbalances on price returns.

First, we revisit the significance of contemporaneous cross-impact by considering various definitions of order flow imbalance (OFI). Instead of only looking at the best-level orders, we systematically examine the impact of \textbf{multi-level} order flows in a \textbf{cross-asset} setting. 
Our results show that, once   information from multi-level order flow is incorporated in the definition of order flow imbalance, cross-impact terms do not provide additional explanatory power for contemporaneous impact, compared to a parsimonious model without   cross-impact. To the best of our knowledge, this is the first study to comprehensively analyze the relations between contemporaneous individual returns and multi-level orders in both single-asset and multi-asset settings.  

Furthermore, we consider the associated forecasting problem and investigate the predictive power of the cross-asset order flows on future price returns. Our results suggest that cross-impact terms do provide significant information content for intraday forecasting of future returns over a short horizon of up to several minutes, but their predictability decays quickly through time. 

\subsection{Outline}  Section \ref{sec_ci:notation} describes our dataset and defines the variables of interest. Section \ref{sec_ci:ci} discusses modeling of contemporaneous cross-impact. In Section \ref{sec_ci:forward}, we first discuss the out-of-sample forecasting performance of cross-impact models over one-minute-ahead horizon from two perspectives: $R^2$ values and economic gains, and then examine the predictability over longer horizons. Finally, we conclude the analysis in Section \ref{sec_ci:conclusion} and {highlight potential future research directions}.

\section{Data and variables} \label{sec_ci:notation}
\subsection{Data} \label{sec_ci:data}
We use the Nasdaq ITCH data from LOBSTER to compute the independent and dependent variables. Our data includes the top 100 components of S\&P 500 index, existing from 2017-01-01 to 2019-12-31.\footnote{We select the top 100 components based on their market capitalization as of the most recent market close at the time of our analysis, i.e. 2019-12-31.} 

\citet{cont2014price} found that over short time intervals, price changes are mainly driven by the Order Flow Imbalance (henceforth denoted as OFI).  \citet{kolm2021deep} also demonstrated that forecasting deep learning models trained on OFIs significantly outperform most models trained directly on order books or returns. Therefore, we adopt the OFIs as features in our below analysis.

{During the interval $(t-h, t]$, we enumerate the observations of all order book updates by $n$. Given two consecutive order book states for a given stock $i$ at $n-1$ and $n$, we compute the bid order flows ($\mathrm{OF}_{i, n}^{m, b}$) and ask order flows ($\mathrm{OF}_{i, n}^{m, a}$) of stock $i$ at level $m$ at time $n$ as}

\begin{equation}
\begin{array}{l}
\mathrm{OF}_{i, n}^{m, b}:=\left\{\begin{array}{ll}
q_{i, n}^{m, b}, & \text { if } P_{i, n}^{m, b}>P_{i, n-1}^{m, b}, \\\nonumber
q_{i, n}^{m, b}-q_{i, n-1}^{m, b}, & \text { if } P_{i, n}^{m, b}=P_{i, n-1}^{m, b}, \\
-q_{i, n}^{m, b}, & \text { if } P_{i, n}^{m, b}<P_{i, n-1}^{m, b},
\end{array}\right. \\
\end{array}
\end{equation}

\begin{equation}
\begin{array}{l}
\mathrm{OF}_{i, n}^{m, a}:=\left\{\begin{array}{ll}
-q_{i, n}^{m, a}, & \text { if } P_{i, n}^{m, a}>P_{i, n-1}^{m, a}, \\\nonumber
q_{i, n}^{m, a}-q_{i, n-1}^{m, a}, & \text { if } P_{i, n}^{m, a}=P_{i, n-1}^{m, a} \\
q_{i, n}^{m, a}, & \text { if } P_{i, n}^{m, a}<P_{i, n-1}^{m, a},
\end{array}\right.
\end{array}
\end{equation}
{where, $P_{i, n}^{m, b}$ and $q_{i, n}^{m, b}$ denote the bid price and size (in number of shares) of stock $i$ at level $m$, respectively. Similarly, $P_{i, n}^{m, a}$ and $q_{i, n}^{m, a}$ denote the ask price and ask size at level $m$, respectively. Note that the variable $\text{OF}_{i, t}^{m, b}$ is positive when (i) the bid price increase; (ii) the bid price remains the same and the bid size increases. $\text{OF}_{i, t}^{m, b}$ is negative when (i) the bid price decreases; (ii) the bid price remains the same and the bid size decreases. One can perform an analogous analysis and interpretation for the ask order flows $\text{OF}_{i, t}^{m, a}$.}

\paragraph{Best-level OFI.} It calculates the accumulative OFIs at {the best bid/ask side} during a given time interval (see  \citet{cont2014price, kolm2021deep}), and is defined as\footnote{{In \citet{cont2014price}, OFI was mathematically defined as $\text{OFI}_{i, t}^{1, h} = L_{i, h}^{1, b}  - C_{i, h}^{1, b} - M_{i, h}^{1, b} - L_{i, h}^{1, a} + C_{i, h}^{1, a} - M_{i, h}^{1, a}$, where $L_{i, h}^{1, b}$ denotes the total size of buy orders that arrived to the current best bid during the time interval $(t-h, t]$; $C_{i, h}^{1, b}$ denotes the total size of buy orders that canceled from the current best bid during the time interval $(t-h, t]$; $M_{i, h}^{1, b}$ denotes the total size of marketable buy orders that arrived to current best ask during the time interval $(t-h, t]$. The quantities $L_{i, h}^{1, a}, C_{i, h}^{1, a}, M_{i, h}^{1, a}$ for sell orders are defined analogously. However, in the empirical study of \citet{cont2014price}, the OFI was computed from fluctuations in best bid/ask prices and their sizes according to Eqn \eqref{eq:OFI}. The reason is that information about individual orders is not available in the data set. For better comparison, we employ the same formula, i.e. Eqn \eqref{eq:OFI}, to compute OFI.}}
\begin{equation}\label{eq:OFI}
\text{OFI}_{i, t}^{1, h} \coloneqq \sum_{n = N(t-h)+1}^{N(t)} \mathrm{OF}_{i, n}^{1, b} - \mathrm{OF}_{i, n}^{1, a},
\end{equation}
where $N(t-h)+1$ and $N(t)$ are the indexes of the first and the last order book event in the interval $(t-h, t]$.

\paragraph{Deeper-level OFI.} 
A natural extension of the best-level OFI defined in Eqn \eqref{eq:OFI} is deeper-level OFI (see \citet{xu2018multi, kolm2021deep}). We define OFI at level $m$ ($m \geq 1$) as follows 
\begin{equation} \label{eq:def_mein}
\text{OFI}_{i, t}^{m, h} \coloneqq \sum_{n = N(t-h)+1}^{N(t)} \mathrm{OF}_{i, n}^{m, b} - \mathrm{OF}_{i, n}^{m, a}.
\end{equation}

Due to the intraday pattern in limit order depth, we use the average size to scale OFIs at the corresponding levels (consistent with \citet{ahn2001limit, harris2005information}), and consider 
\begin{equation}
\text{ofi}_{i, t}^{m, h} = \frac{\mathrm{OFI}_{i, t}^{m, h}}{Q^{M, h}_{i, t}},
\end{equation}
where $Q^{M, h}_{i, t} = \frac{1}{M} \sum_{m=1}^M \frac{1}{2 \Delta N(t)} \sum_{n=N(t-h)+1}^{N(t)} \left[ q_{i, n}^{m, b} +  q_{i, n}^{m, a}\right]$ is the average order book depth across the first $M$ levels and $\Delta N(t) = N(t)-N(t-h)$ is the number of events during $(t-h, t]$. In this paper, we consider the top $M=10$ levels of LOB and denote the multi-level OFI vector as $\pmb{\mathrm{ofi}}_{i, t}^{(h)} = \left(\mathrm{ofi}_{i, t}^{1, h}, \cdots, \mathrm{ofi}_{i, t}^{10, h}\right)^{T}$.

\paragraph{Integrated OFI.} 
Our following analysis in Section \ref{sec_ci:sum_stat} will show that there exist strong correlations between multi-level OFIs, and that the first principal component can explain over 89\% of the total variance among multi-level OFIs. In order to make use of the information embedded in multiple LOB levels and avoid overfitting, we propose an integrated version of OFIs via Principal Components Analysis (PCA)
as shown in Eqn \eqref{eq:int_ofi}, which only preserves the first principal component.\footnote{We would like to thank an anonymous reviewer for suggesting an analysis of various aggregations of multi-level OFIs, as detailed in Appendix \ref{sec_ci:integrate}.} We further normalize the first principal component by dividing by its $l_1$ norm so that the weights of multi-level OFIs in constructing integrated OFIs sum to 1, leading to 
\begin{equation}\label{eq:int_ofi}
\text{ofi}_{i, t}^{I, h} = \frac{\pmb{w}_1^T\pmb{\mathrm{ofi}}_{i, t}^{(h)}}{\|\pmb{w}_1\|_1},
\end{equation} 
where $\pmb{w}_1$ is the first principal vector computed from historical data. To the best of our knowledge, this is the first work to {{aggregate}  multi-level OFIs into a single variable}.

  
\paragraph{Logarithmic returns.} Our dependent variable is the logarithmic asset return. Specifically, we define the returns over the interval $(t-h, t]$ as follows:
\vspace{-1mm}
\begin{equation}
r_{i, t}^{(h)}=\log \left(\frac{P_{i, t}}{P_{i, t-h}}\right),
\end{equation}
where $P_{i, t}$ is the mid-price at time $t$, i.e. $P_{i, t} = \frac{P_{i, t}^{1, b} + P_{i, t}^{1, a}}{2}$.

\subsection{Summary statistics}\label{sec_ci:sum_stat}
Table \ref{tab:data} presents the summary statistics of multi-level OFIs, integrated OFIs, and returns for the top 100 components of S\&P 500 index. These descriptive statistics (e.g. mean, std, etc) are computed at the minute level and aggregated across trading days and stocks.

\begin{figure}[H]
    \centering
\subfigure[Average]{
\includegraphics[width=.22\textwidth, trim=2.3cm 1.8cm 2.5cm 2.5cm,clip]{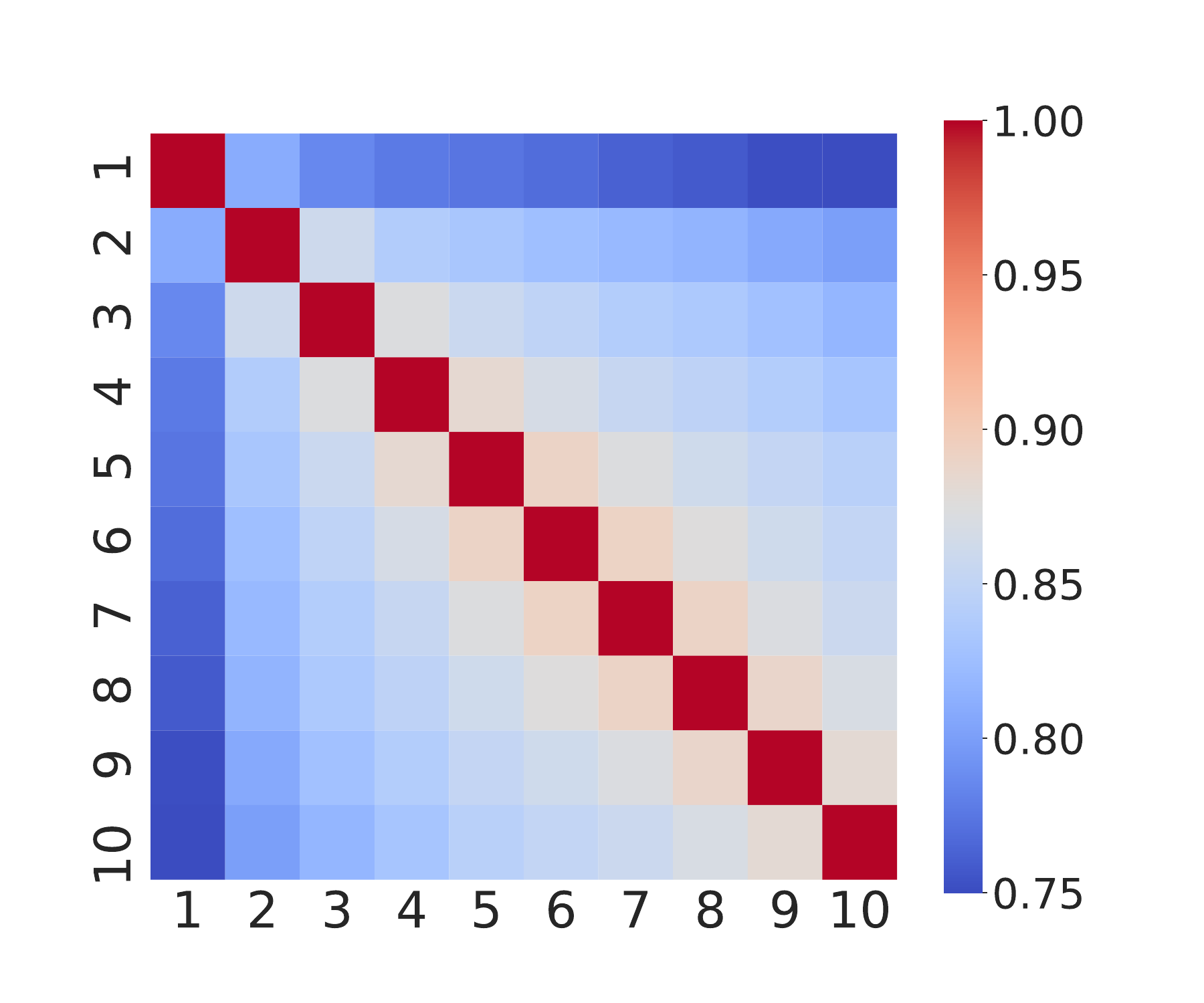}
}
\subfigure[AAPL]{
\includegraphics[width=.22\textwidth, trim=2.3cm 1.8cm 2.5cm 2.5cm,clip]{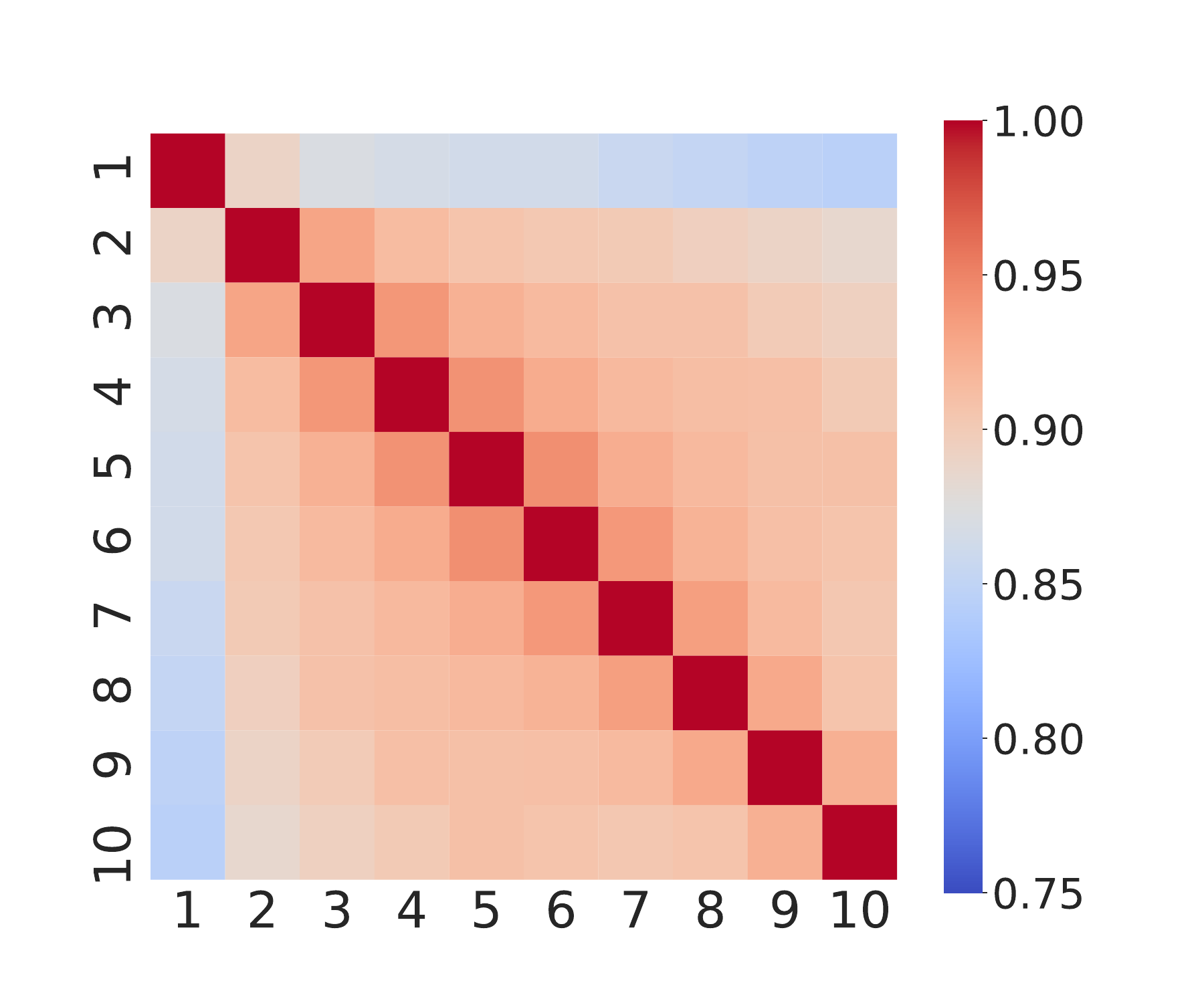}
}
\subfigure[JPM]{
\includegraphics[width=.22\textwidth, trim=2.3cm 1.8cm 2.5cm 2.5cm,clip]{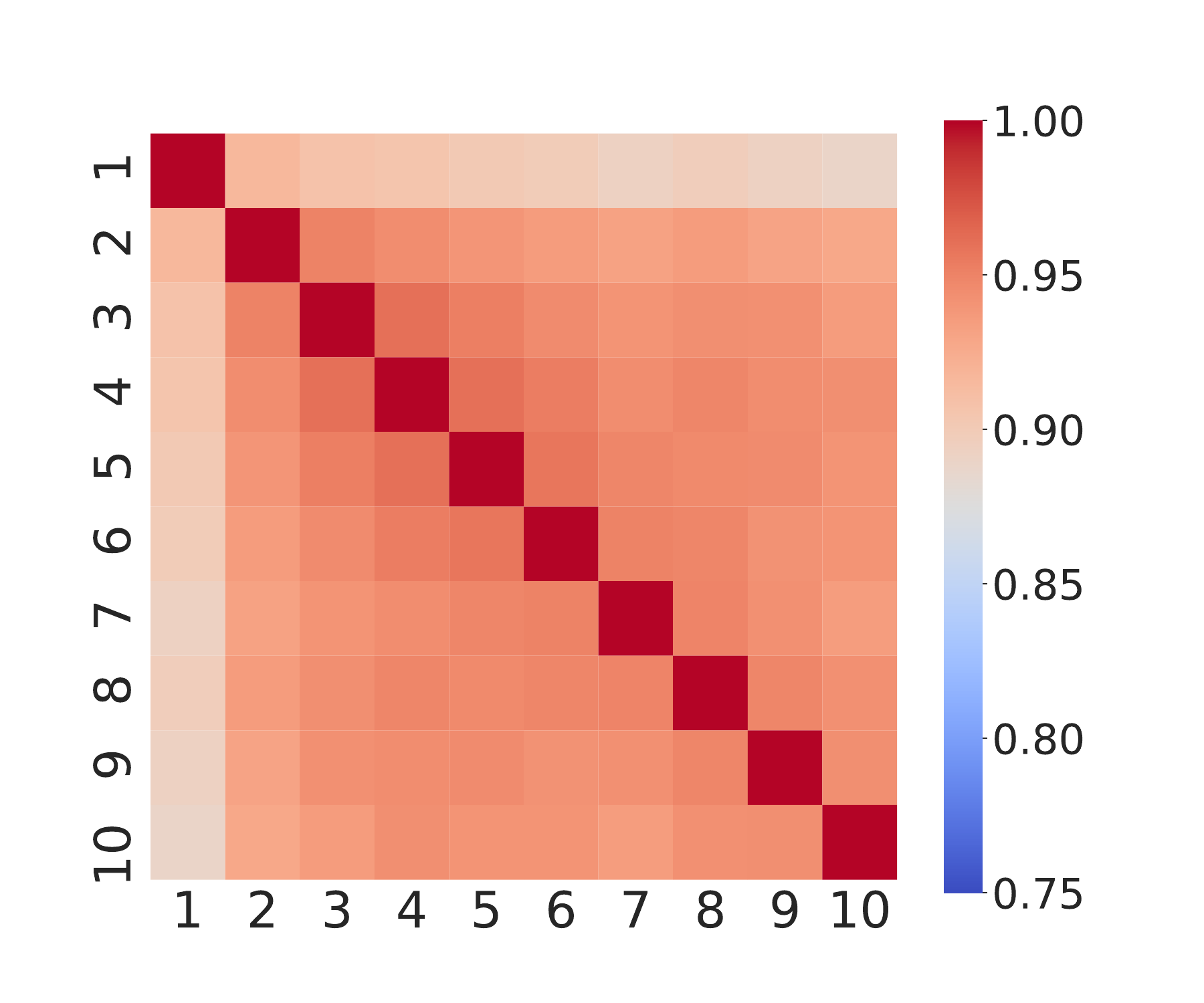}
}
\subfigure[JNJ]{
\includegraphics[width=.22\textwidth, trim=2.3cm 1.8cm 2.5cm 2.5cm,clip]{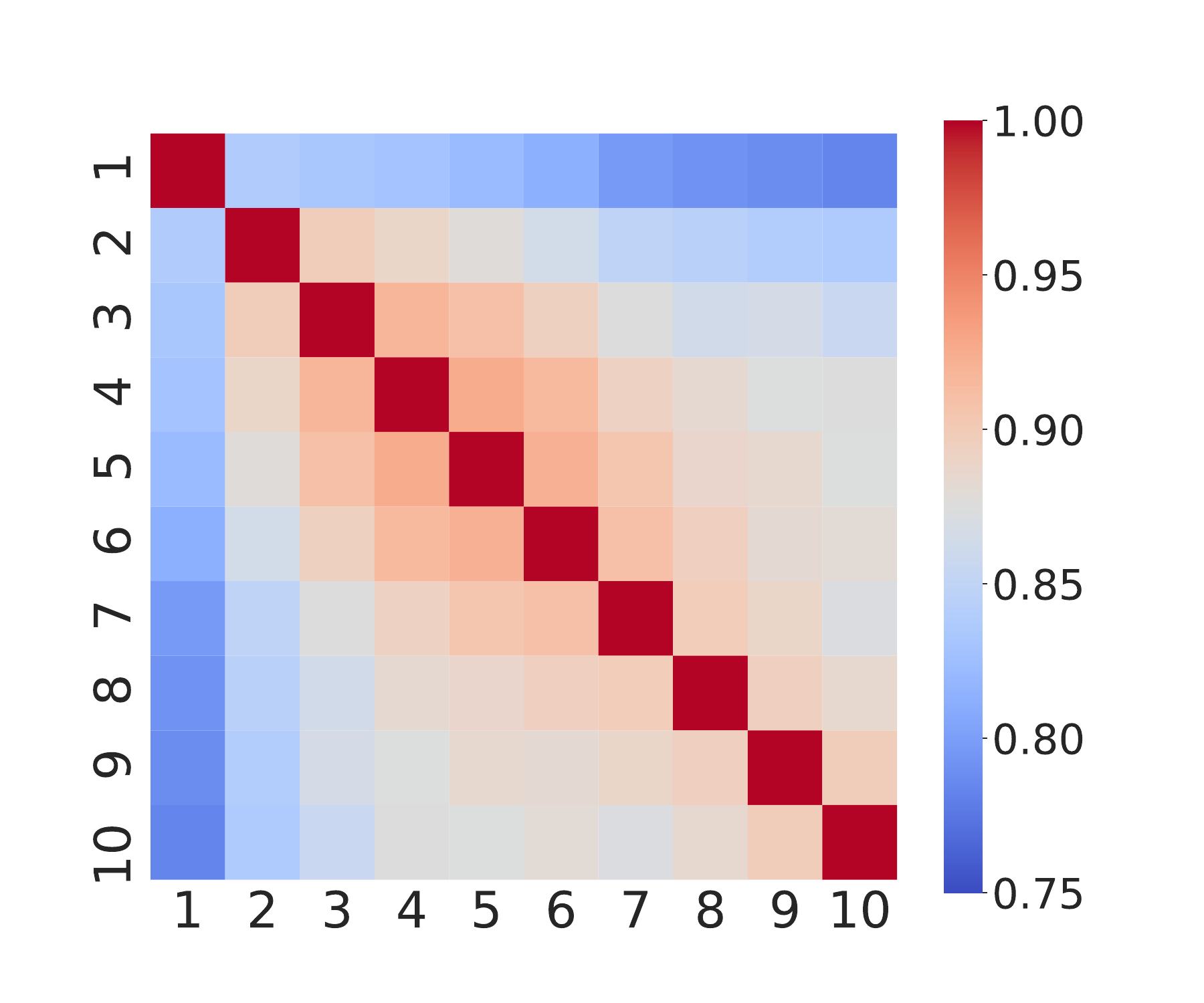}
}
\caption{Correlation matrix of multi-level OFIs. }
\caption*{\textit{Note:} Plot (a) is averaged across each stock and each trading day, Plots (b)-(d): correlation matrix of Apple (AAPL), JPMorgan Chase (JPM), and Johnson \& Johnson (JNJ) averaged across each trading day. The $x$-axis and $y$-axis represent different levels of OFIs.}
\label{fig:corr}
\end{figure}

Figure \ref{fig:corr} reveals that even though the correlation structure of multi-level OFIs may vary across stocks, they all show strong relationships (above 75\%). It is worth pointing out that the best-level OFI exhibits the smallest correlation with any of the remaining nine levels, a pattern that persists across different stocks. Table \ref{tab:evr_ofi_multi} further reveals that the first principal component explains more than 89\% of the total variance. 

\begin{table}[H]
    \centering
    \caption{Summary statistics of OFIs and returns.} 
    \resizebox{1.0\textwidth}{!}{\begin{tabular}{lccccccccc}
\toprule
 & Mean (bp)      & Std (bp)     & Skewness      & Kurtosis   & 10\% (bp)  & 25\% (bp)  & 50\% (bp) & 75\% (bp)  & 90\% (bp) \\
\midrule
$\text{ofi}^{1, (1m)}$  & -0.01 & 6.26  & -0.04 & 1.89 & -7.97  & -3.45  & 0.03 & 3.47  & 7.90  \\
$\text{ofi}^{2, (1m)}$  & 0.01  & 6.86  & -0.04 & 1.04 & -8.86  & -3.88  & 0.02 & 3.95  & 8.85  \\
$\text{ofi}^{3, (1m)}$  & -0.01 & 7.05  & -0.04 & 0.71 & -9.26  & -4.08  & 0.01 & 4.11  & 9.19  \\
$\text{ofi}^{4, (1m)}$  & -0.02 & 7.22  & -0.05 & 0.68 & -9.50  & -4.21  & 0.01 & 4.24  & 9.40  \\
$\text{ofi}^{5, (1m)}$  & -0.03 & 7.14  & -0.05 & 0.79 & -9.38  & -4.14  & 0.01 & 4.15  & 9.25  \\
$\text{ofi}^{6, (1m)}$  & -0.03 & 6.87  & -0.04 & 0.96 & -8.98  & -3.94  & 0.01 & 3.95  & 8.85  \\
$\text{ofi}^{7, (1m)}$  & -0.03 & 6.39  & -0.05 & 1.29 & -8.31  & -3.59  & 0.01 & 3.59  & 8.16  \\
$\text{ofi}^{8, (1m)}$  & -0.03 & 6.03  & -0.05 & 1.59 & -7.80  & -3.37  & 0.01 & 3.36  & 7.66  \\
$\text{ofi}^{9, (1m)}$  & -0.05 & 5.71  & -0.05 & 1.96 & -7.38  & -3.18  & 0.01 & 3.14  & 7.19  \\
$\text{ofi}^{10, (1m)}$ & -0.05 & 5.38  & -0.05 & 2.52 & -6.92  & -2.97  & 0.01 & 2.91  & 6.74  \\
$\text{ofi}^{I, (1m)}$  & 0.01  & 6.53  & -0.05 & 0.76 & -8.52  & -3.81  & 0.05 & 3.89  & 8.47  \\\midrule
$r^{(1m)}$              & 0.02  & 4.81 & -0.04 & 1.85 & -6.22 & -2.71 & 0.00 & 2.79 & 6.23 \\
\bottomrule
\end{tabular}}
    \caption*{\textit{Note:} These statistics are computed at the minute level across each stock and the full sample period. 1bp = 0.0001 = 0.01\%.}
    \label{tab:data}
\end{table}

\begin{table}[H]
    \centering
    \caption{Average percentage and the standard deviation (in parentheses) of variance attributed to each principal component.} 
    \resizebox{1.0\textwidth}{!}{\begin{tabular}{lcccccccccc}\toprule
Principal Component & 1      & 2   & 3  & 4  & 5  & 6  & 7  & 8  & 9  & 10  \\\midrule
Explained Variance Ratio & 89.06 & 4.99 & 2.28 & 1.28 & 0.80 & 0.54 & 0.39 & 0.29 & 0.21 & 0.15 \\
 & (6.12) & (3.52) & (1.26) & (0.74) & (0.48) & (0.34) & (0.25) & (0.19) & (0.15) & (0.11) \\\bottomrule

\end{tabular}
}
    \caption*{\textit{Note:} The table reports the ratio (in percentage points) between the variance of each principal component and the total variance averaged across each stock and trading day.}
    \label{tab:evr_ofi_multi}
\end{table}

\begin{figure}[H]
    \centering
\subfigure[Average\label{fig:PC1_OFI_aver}]{
\includegraphics[width=.22\textwidth, trim=5mm 1cm 2.2cm 1.5cm,clip]{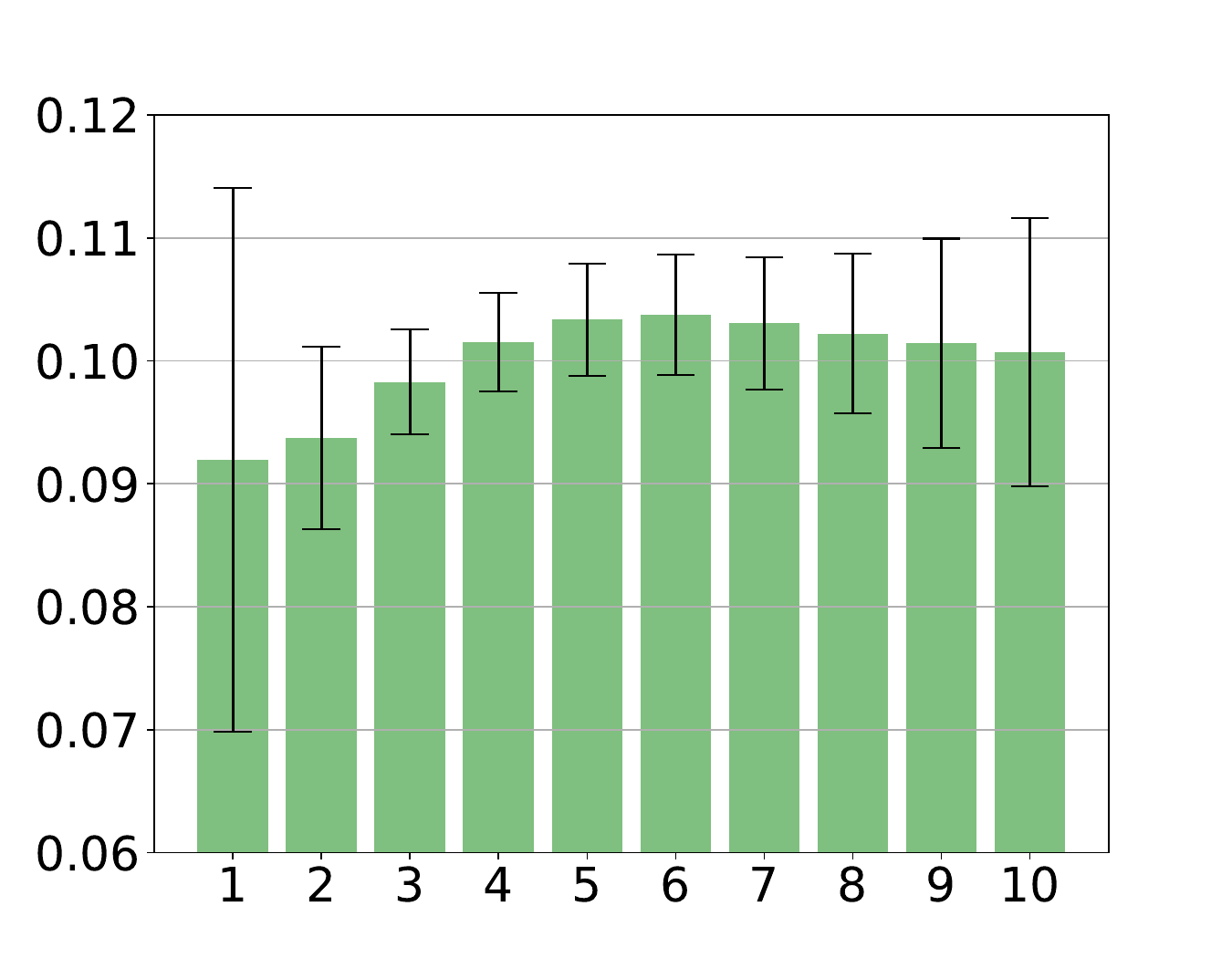}
}
\subfigure[Volume\label{fig:PC1_OFI_volume}]{
\includegraphics[width=.22\textwidth, trim=5mm 1cm 2.2cm 1.5cm,clip]{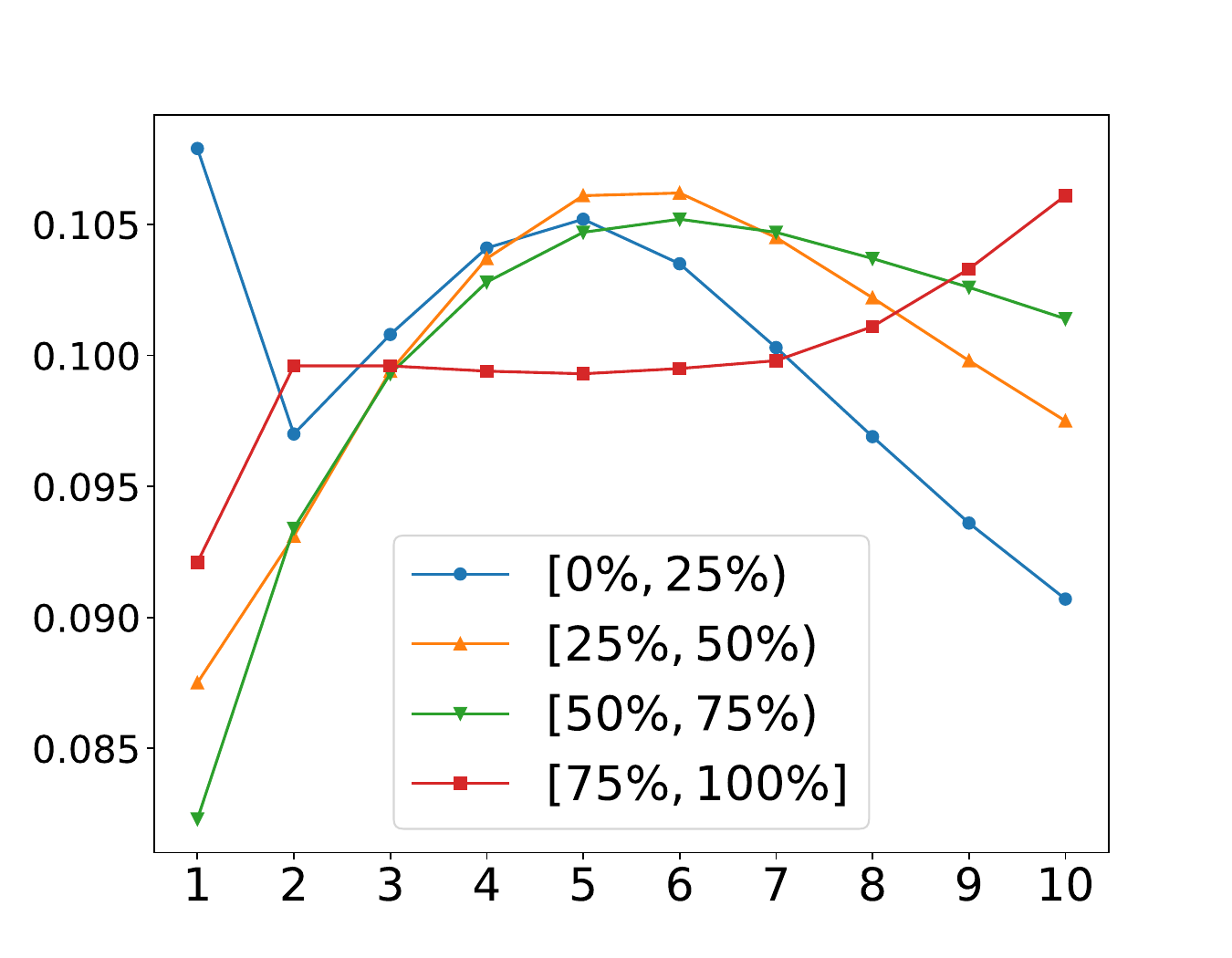}
}
\subfigure[Volatility\label{fig:PC1_OFI_volatility}]{
\includegraphics[width=.22\textwidth, trim=5mm 1cm 2.2cm 1.5cm,clip]{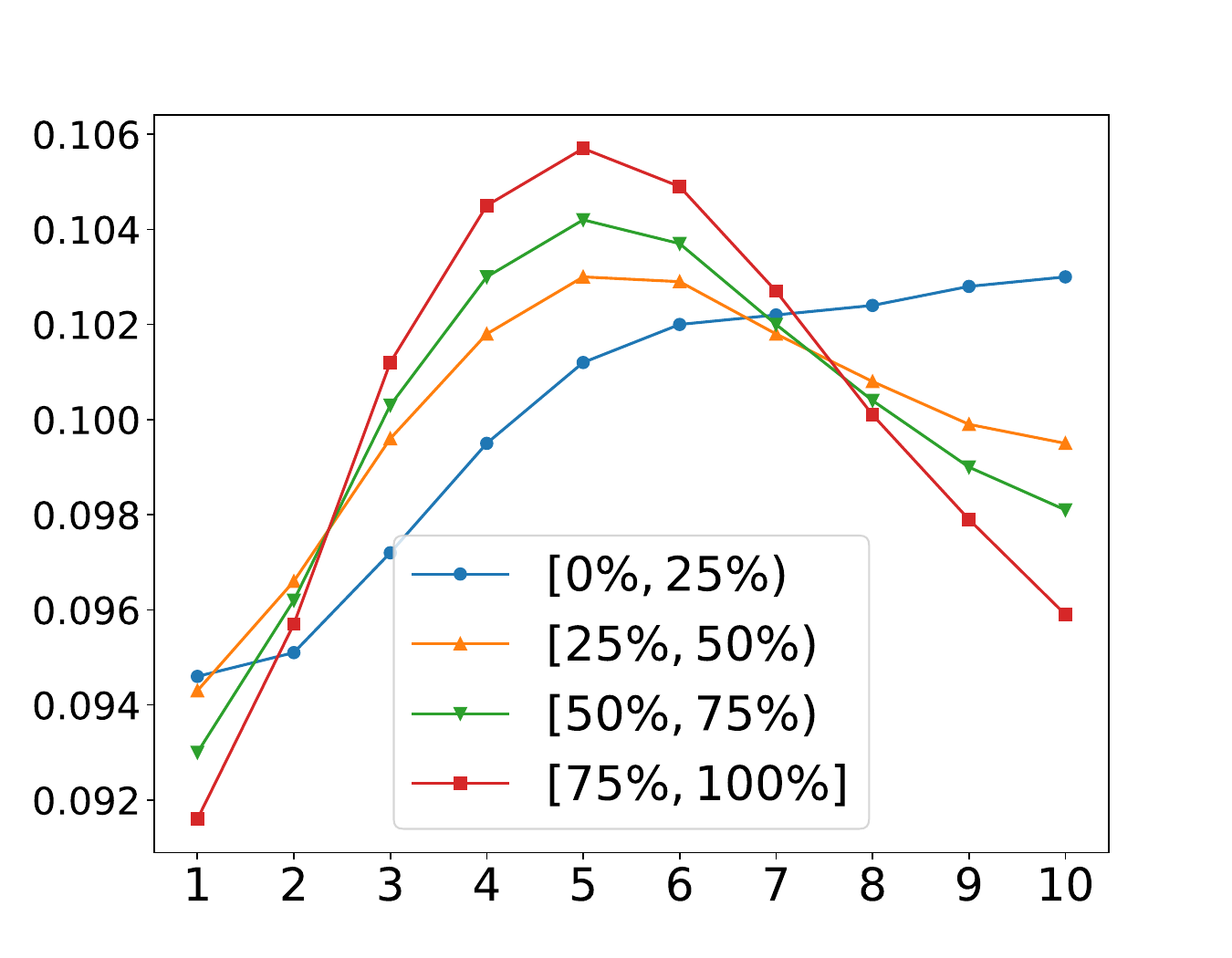}
}
\subfigure[Spread\label{fig:PC1_OFI_spread}]{
\includegraphics[width=.22\textwidth, trim=5mm 1cm 2.2cm 1.5cm,clip]{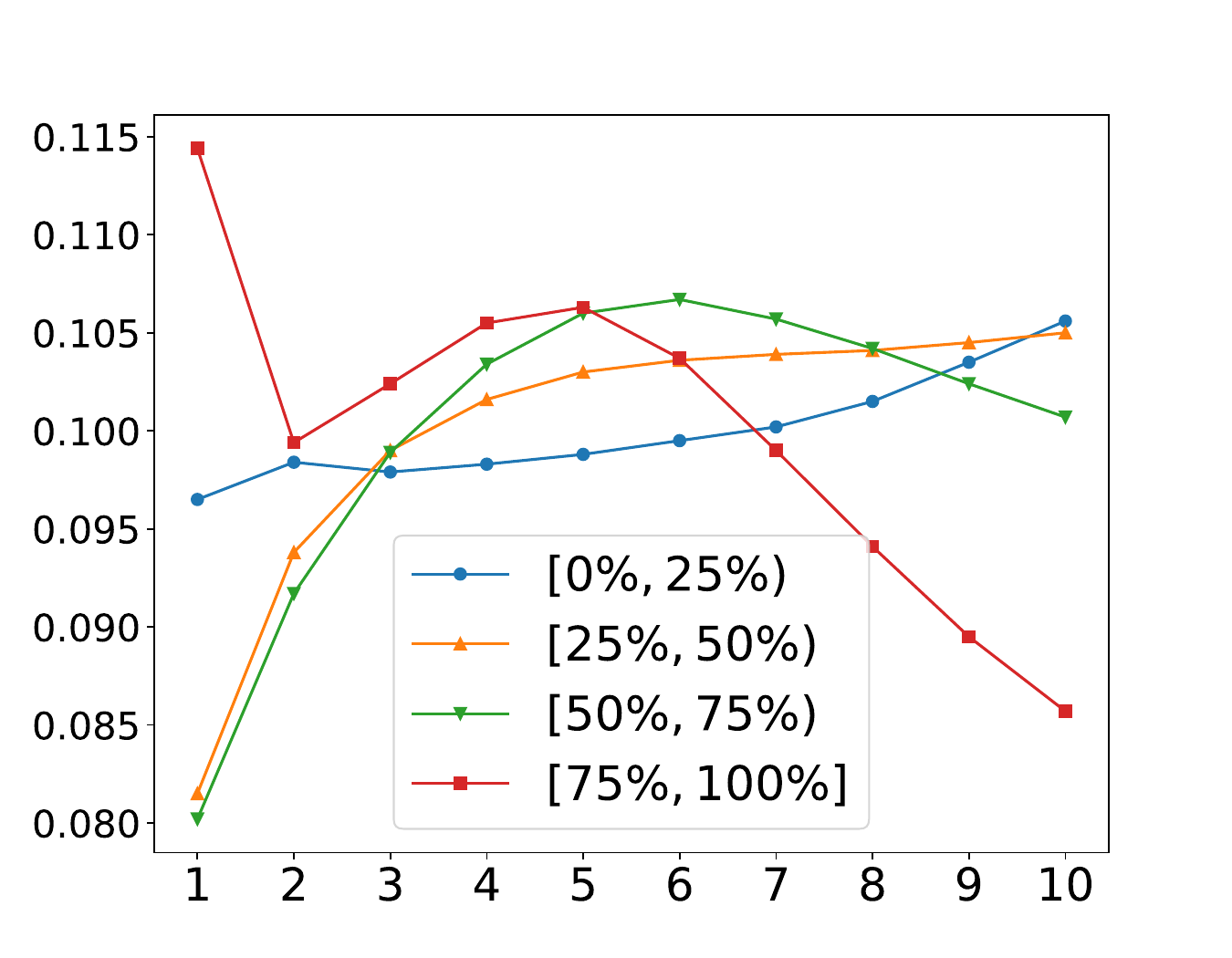}
}
\caption{First principal component of multi-level OFIs, in quantile buckets for various stock characteristics. }
\caption*{\textit{Note:} The $x$-axis indexes the top 10 levels of the OFIs. {Volume}: trading volume on the previous trading day. {Volatility}: volatility of 1-minute returns during the previous trading day. {Spread}: average bid-ask spread during the previous trading day.  $[0\%, 25\%)$, respectively $[75\%, 100\%]$, denote the subset of stocks with the lowest, respectively highest, 25\% values for a given stock characteristic.}
\label{fig:PC1_OFI_detail}
\end{figure}

In Figure \ref{fig:PC1_OFI_detail}, we show statistics pertaining to the weights attributed to the top 10 levels in the first principal component. Plot \ref{fig:PC1_OFI_aver} shows the average weights, and the one standard deviation bars, across all stocks in the universe. Plot \ref{fig:PC1_OFI_aver} reveals that the best-level OFI has the smallest weight in the first principal component, but the highest standard deviation, hinting that it fluctuates significantly across stocks. Plots (b-d) show various patterns for the first principal component of multi-level OFIs, for each quantile bucket of various stock characteristics, in particular, for volume, volatility and spread. For instance, in Figure \ref{fig:PC1_OFI_volume},  the red curve shows the average weights in the first principal component for each of the 10 levels, where the average is taken over all the top 25\% largest volume stocks. A striking pattern that emerges from this figure is that for \textit{high-volume} (red line in \ref{fig:PC1_OFI_volume}), and \textit{low-volatility stocks} (blue line in \ref{fig:PC1_OFI_volatility}), OFIs deeper in the LOB receive more weights in the first component. However, for \textit{low-volume} (blue line in \ref{fig:PC1_OFI_volume}), and \textit{large-spread stocks} (red line in \ref{fig:PC1_OFI_spread}), the best-level OFIs account more than the deeper-level OFIs.

\section{Contemporaneous cross-impact} \label{sec_ci:ci}
In this section, we study the existence of contemporaneous cross-impact by comparing it with the price impact model studied in \citet{cont2014price}. 


\subsection{Models}
\paragraph{Price impact of best-level OFIs.} We first pay attention to the price impact of best-level OFI ($\mathrm{ofi}_{i, t}^{1, h}$) on {contemporaneous returns} ($r_{i, t}^{(h)}$) that materialize over the same time bucket as the OFI, via the model 
\begin{equation}\label{eq:pi_best}
\textbf{PI}^{[1]}:\quad r_{i, t}^{(h)} = \alpha_{i}^{[1]} + \beta_{i}^{[1]} \mathrm{ofi}_{i, t}^{1, h}+\epsilon_{i, t}^{[1]}.
\end{equation} 
\noindent Here, $\alpha_{i}^{[1]}$ and $\beta_{i}^{[1]}$ are the intercept and slope coefficients, respectively. $\epsilon_{i, t}^{[1]}$ is a noise term summarizing the influences of other factors, such as the OFIs at even deeper levels, and potentially the trading behaviors of other stocks. For the sake of simplicity, we refer to the above regression model as $\textbf{PI}^{[1]}$ and use ordinary least squares (OLS) to estimate it.

\paragraph{Price impact of integrated OFIs.} The second model specification takes into account the impact of multi-level OFIs by leveraging the integrated OFIs, which we set up as follows and use OLS for estimation.
\begin{equation}\label{eq:pi_int}
\textbf{PI}^{I}:\quad r_{i, t}^{(h)} = \alpha_{i}^{I} + \beta_{i}^{I} \text{ofi}_{i, t}^{{I}, h} +\epsilon_{i, t}^{I}.
\end{equation}

\paragraph{Cross-impact of best-level OFIs.} 
Assuming there are $N$ stocks in the studied universe, we incorporate the multi-asset best-level OFIs, $\text{ofi}_{j, t}^{1, h} (j=1, \dots, N)$,  as candidate features to help fit the returns of the $i$-th stock $r_{i, t}^{(h)}$. For simplicity, we denote the impact from itself (stock $i$) as \textit{Self} and that from other stocks as \textit{Cross}, as shown below,
\begin{equation}
\label{eq:ci_best}
\textbf{CI}^{[1]}:\quad r_{i, t}^{(h)}  = \alpha_{i}^{[1]} +  \underbrace{\beta_{i, i}^{[1]} \text{ofi}_{i, t}^{1, h}}_\textit{Self} + \sum_{j\neq i} \underbrace{\beta_{i, j}^{[1]} \text{ofi}_{j, t}^{1, h}}_\textit{Cross} + \eta_{i, t}^{[1]}.
\end{equation}
Therefore, $\beta_{i, j}^{[1]}$ represents the influence of the $j$-th stock's best-level OFIs on the returns of stock $i$.

\paragraph{Cross-impact of integrated OFIs.} Finally, we incorporate the cross-asset integrated OFIs to explore the impact of multi-level OFIs from other assets, resulting in the following $\textbf{CI}^{I}$ model,
\begin{equation}
\label{eq:ci_int}
\textbf{CI}^{I}:\quad r_{i, t}^{(h)}  = \alpha_{i}^{I} + \underbrace{\beta_{i,i}^{I} \mathrm{ofi}_{i, t}^{I, h}}_\textit{Self}+ \sum_{j \neq i} \underbrace{\beta_{i, j}^{I} \text{ofi}_{j, t}^{I, h}}_\textit{Cross} + \eta_{i, t}^{I}.
\end{equation}

\begin{figure}[H]
    \centering
    \includegraphics[width=.56\textwidth, trim=1cm 1cm 2cm 1cm,clip]{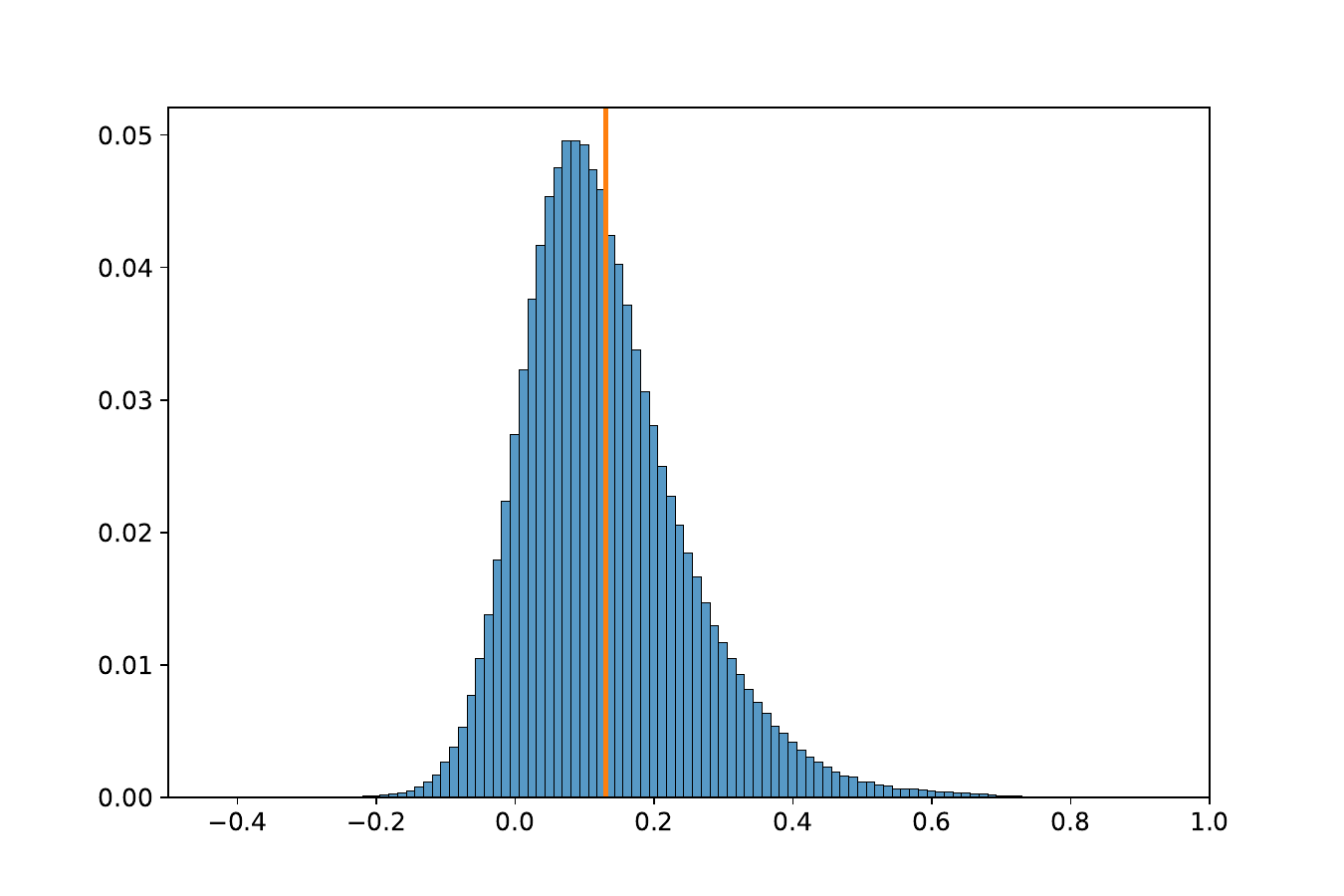}
    \vspace{-2mm}
\caption{Distribution of correlations based on the best-level OFIs. }
\caption*{\textit{Note:} The orange vertical line represents the average correlation.}
\label{fig:cx_corr}
\end{figure}

\paragraph{Sparsity of cross-impact.} As we are aware, OLS regression becomes ill-posed when there are fewer observations than parameters. Recall that we are now considering $N \approx 100$ independent variables in Eqns \eqref{eq:ci_best} and \eqref{eq:ci_int}. Assuming the time interval is one minute and we are interested in estimating the intraday cross-impact models, e.g. relying on the 30-min estimation window and 1-min returns (as in \citet{cont2014price}), then it seems inappropriate to estimate $\textbf{CI}^{[1]}$ and $\textbf{CI}^{I}$ for intraday scenarios using the OLS regression with more variables than observations. Moreover, the multicollinearity issue of features contradicts the necessary condition for a well-posed OLS. As displayed in Figure \ref{fig:cx_corr}, a significant portion of the cross-asset correlations based on the best-level OFIs cannot be ignored. For example, approximately 10\% of correlations are larger than 0.30.   
Last, \citet{capponi2020multi} found that a certain number of cross-impact coefficients from their OLS regressions are not statistically significant at the 1\% significance level.

With all the above considerations in mind, we assume that there is a small number of assets having a significant impact on a specific stock $i$, as opposed to the entire universe, in $\textbf{CI}^{[1]}$ and $\textbf{CI}^{I}$. To this end, we {apply the Least Absolute Shrinkage and Selection Operator (LASSO)\footnote{LASSO is a regression method that performs both variable selection and regularization, in order to enhance the prediction accuracy and interpretability of regression models (see more in \citet{hastie2009elements}). It can be formulated as a linear regression model and the objective function consists of two parts, i.e. the sum of squared residuals, and the $l_1$ constraint on the regression coefficients. In this work, we employ the cross-validation to choose the $l_1$ penalty weight for each regression.} to solve Eqns \eqref{eq:ci_best} and \eqref{eq:ci_int}}. The sparsity of cross-impact terms also facilitates the explanation of coefficients. Note that even though the sparsity of the cross-impact terms is not theoretically guaranteed, our empirical evidence confirms this modeling assumption. 

\subsection{Empirical results}\label{sec_ci:pi_exp}
For a more representative and fair comparison with previous studies, we apply a similar procedure described in \citet{cont2014price} to our experiments. We exclude the first and last 30 minutes of the trading day due to the increased volatility near the opening and closing sessions, in line with \citet{hasbrouck2002limit, chordia2002order, chordia2004order, cont2014price, capponi2020multi}. In particular, we use each non-overlapping 30-minute estimation window during the intraday time interval 10:00 am - 3:30 pm to estimate the regressions, namely Eqns \eqref{eq:pi_best}, \eqref{eq:pi_int}, \eqref{eq:ci_best}, and \eqref{eq:ci_int}. Within each window, returns and OFIs are computed for every minute. 

\subsubsection{In-sample performance} 
We first measure the model performance via in-sample adjusted-$R^2$, denoted as the \textbf{in-sample $R^2$} or \textbf{IS $R^2$}. From Table \ref{tab:pi_ci_is}, we first observe that $\textbf{PI}^{[1]}$ can explain 71.16\% of the in-sample variation of a stock's contemporaneous returns, consistent with the findings of \citet{cont2014price}. Meanwhile, $\textbf{PI}^{I}$ displays higher and more consistent explanation power, with an average adjusted $R^2$ value of 87.14\% and a standard deviation of 9.16\%, indicating the effectiveness of our integrated OFIs.\footnote{We also investigate price impact with multi-level OFIs in Appendix \ref{sec_ci:pi_multi}. The results demonstrate that the price impact model using integrated OFIs outperforms those using multi-level OFIs in out-of-sample tests.}

Table \ref{tab:pi_ci_is} also shows that the in-sample $R^2$ values increase as cross-asset OFIs are included as additional features, which is not surprising given that $\textbf{PI}^{[1]}$ (respectively, $\textbf{PI}^{I}$) is a nested model of $\textbf{CI}^{[1]}$ (respectively, $\textbf{CI}^{I}$). However, the increments of the in-sample $R^2$ are smaller when using integrated OFIs (87.85\%-87.14\%=0.71\%), compared to the counterpart using best-level OFIs (73.87\%-71.16\%=2.71\%). This indicates that cross-asset multi-level OFIs may not provide additional information on the variance in returns compared to the price impact model with integrated OFIs.

\begin{table}[H]
    \centering
    \caption{{In-sample} performance for contemporaneous returns.}
    \resizebox{0.5\textwidth}{!}{


\begin{tabular}{lcccc}
    \toprule
    & \multicolumn{2}{c}{{Best-level OFIs}} & \multicolumn{2}{c}{{Integrated OFIs}} \\
     \cmidrule(lr){2-3}\cmidrule(lr){4-5} 
 & $\textbf{PI}^{[1]}$ &  $\textbf{CI}^{[1]}$ & $\textbf{PI}^{I}$ & $\textbf{CI}^{I}$ \\\midrule
IS $R^2$  & 71.16 &  73.87 & 87.14  & \text{87.85} \\
& (13.80) & (12.23) & (9.16) & (8.58) \\
\bottomrule
\end{tabular}}
    \caption*{\textit{Note:} The table reports the mean values and standard deviations (in parentheses) of {in-sample} $R^2$ (in percentage points) of various models when modeling contemporaneous returns. The models include $\textbf{PI}^{[1]}$ (Eqn \eqref{eq:pi_best}), $\textbf{CI}^{[1]}$ (Eqn \eqref{eq:ci_best}), $\textbf{PI}^{I}$ (Eqn \eqref{eq:pi_int}), and $\textbf{CI}^{I}$ (Eqn \eqref{eq:ci_int}). These statistics are averaged across each stock and each regression window.}
    \label{tab:pi_ci_is}
\end{table}

Next, we take a closer look at the cross-impact coefficients based on either the best-level or integrated OFIs, i.e. $\beta_{i, j}^{[1]}$ and $\beta_{i, j}^{I} \,\, (i,j=1, \dots, N)$.  Table \ref{tab:coef_ci} reveals the frequency of self-impact and cross-impact variables selected by LASSO, i.e. the frequency  of $\beta_{i, j}^{[1]} \neq 0$ (respectively, $\beta_{i, j}^{I} \neq 0$). We observe that self-impact variables are consistently chosen in both $\textbf{CI}^{[1]}$ and $\textbf{CI}^{I}$, as found in \citet{cont2014price}. However, another interesting observation is that the frequency of a cross-asset integrated OFI variable selected by $\textbf{CI}^{I}$ is around 1/2 of its counterpart in $\textbf{CI}^{[1]}$. When we turn to the size of the average regression coefficients as shown in Table \ref{tab:coef_ci}, we obtain reasonably consistent results. The self-impact is much higher than the cross-impact in both the $\textbf{CI}^{[1]}$ and $\textbf{CI}^{I}$ models, while the cross-impact coefficients in $\textbf{CI}^{I}$ are about 1/3 in scale of their counterparts in $\textbf{CI}^{[1]}$. {This difference in scale may suggest that the cross-impact terms are less important in the $\textbf{CI}^{I}$ model, however, it is worth noting that even small cross-term coefficients can have a non-negligeable effect {when aggregated at portfolio level}.} 

\begin{table}[H]
    \centering
    \caption{Summary statistics of coefficients in the cross-impact models $\textbf{CI}^{[1]}$ and $\textbf{CI}^{I}$.}
    \resizebox{0.5\textwidth}{!}{\begin{tabular}{lcccc}
    \toprule
    & \multicolumn{2}{c}{Frequency (\%)} & \multicolumn{2}{c}{Magnitude} \\
     \cmidrule(lr){2-3}\cmidrule(lr){4-5} 
 & $\textbf{CI}^{[1]}$ &  $\textbf{CI}^{I}$ & $\textbf{CI}^{[1]}$ & $\textbf{CI}^{I}$ \\\midrule
\textit{Self}  & 99.85 &  99.96 & 1.02  & 1.24 \\
 & (0.34) & (0.18) & (0.31) & (0.34) \\\midrule
\textit{Cross}  & 17.34 & 8.29 & $4.5e^{-3}$  & $1.6e^{-3}$ \\
 & (2.78) & (2.56) & ($1.3e^{-3}$) & ($0.7e^{-3}$) \\
\bottomrule
\end{tabular}}
    \caption*{\textit{Note:} The table is calculated over each stock and each  regression window. The first two columns describe the frequency of \textit{Self} and \textit{Cross} variables chosen by the corresponding model with a standard deviation (in parentheses); The last two columns describe the magnitude of \textit{Self} and \textit{Cross} coefficients in the corresponding model with a standard deviation (in parentheses).}
    \label{tab:coef_ci}
\end{table}

\begin{figure}[H]
    \centering
    \includegraphics[width=1.0\textwidth, trim=1cm 0cm 1cm 2mm,clip]{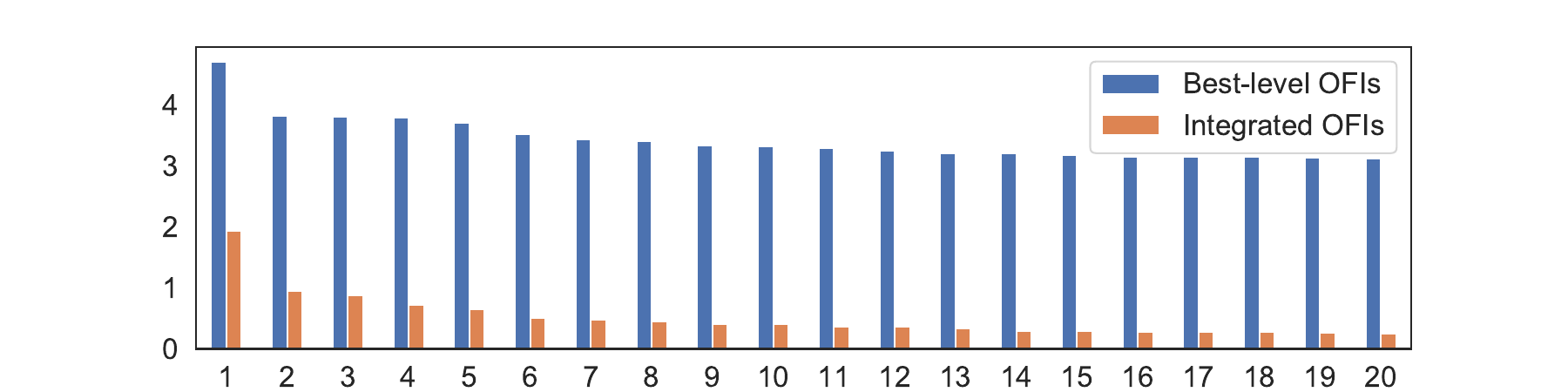}
    \caption{Barplot of singular values for the coefficient matrix in {contemporaneous cross-impact} models.}
    \caption*{\textit{Note:} We perform Singular Value Decomposition (SVD) on the coefficient matrix to obtain the singular values. Singular values are in descending order and the coefficients are averaged over each regression window between 2017–2019. The $x$-axis represents the singular value rank,  and the $y$-axis represents the  singular values.}
    \label{fig:cont_singular}
\end{figure}

In addition, cross-impact being large/small is a statement about a matrix, more related with its singular values and relative magnitudes, rather than the individual value of the coefficients. Figure \ref{fig:cont_singular} shows a comparison of the top 20 singular values of the coefficient matrices given by the best-level and integrated OFIs.\footnote{Here we only use the off-diagonal elements, i.e. $\big[\beta_{i, j}^{[1]} \big]_{i \neq j}$ and $\big[\beta_{i, j}^{I} \big]_{i \neq j}$.} The relatively large singular values of the best-level OFI matrix are a consequence of the higher edge density, and thus average degree, of the network. Note that both networks exhibit a large top singular value of the adjacency matrix (akin to the usual \textit{market mode} in \citet{laloux2000random}), and the integrated OFI network has a faster decay of the spectrum, thus revealing its low-rank structure.

\begin{figure}[H]
    \centering
\subfigure[Threshold = 95-th percentile, based on best-level OFIs \label{fig:freq_lasso_sec}]{
\includegraphics[width=.45\textwidth, trim=3.6cm 4cm 3.6cm 4cm,clip]{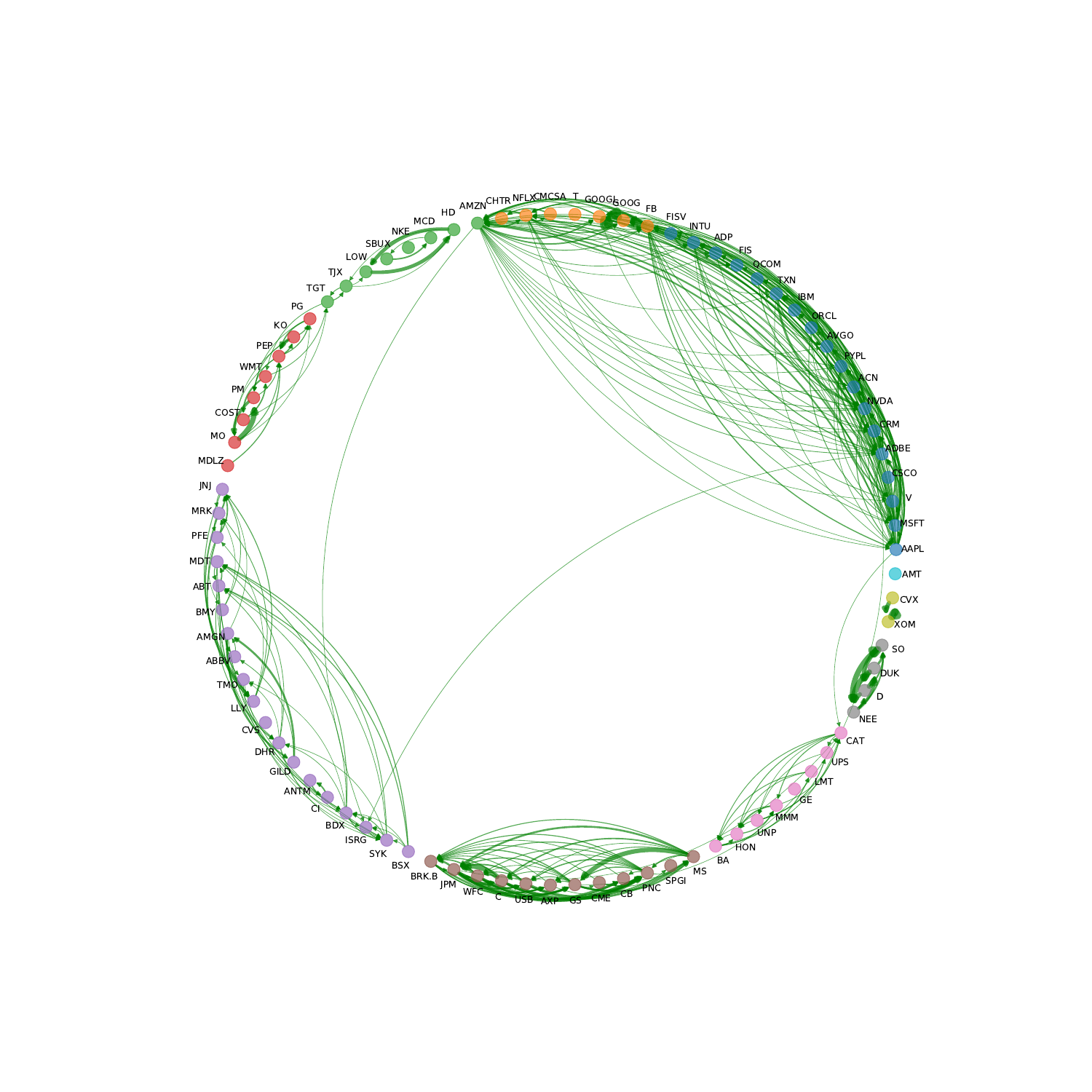}
}
\subfigure[Threshold = 95-th percentile, based on integrated OFIs  \label{fig:freq_lasso_sec_int}]{
\includegraphics[width=.445\textwidth, trim=2.6cm 3cm 2.6cm 3cm,clip]{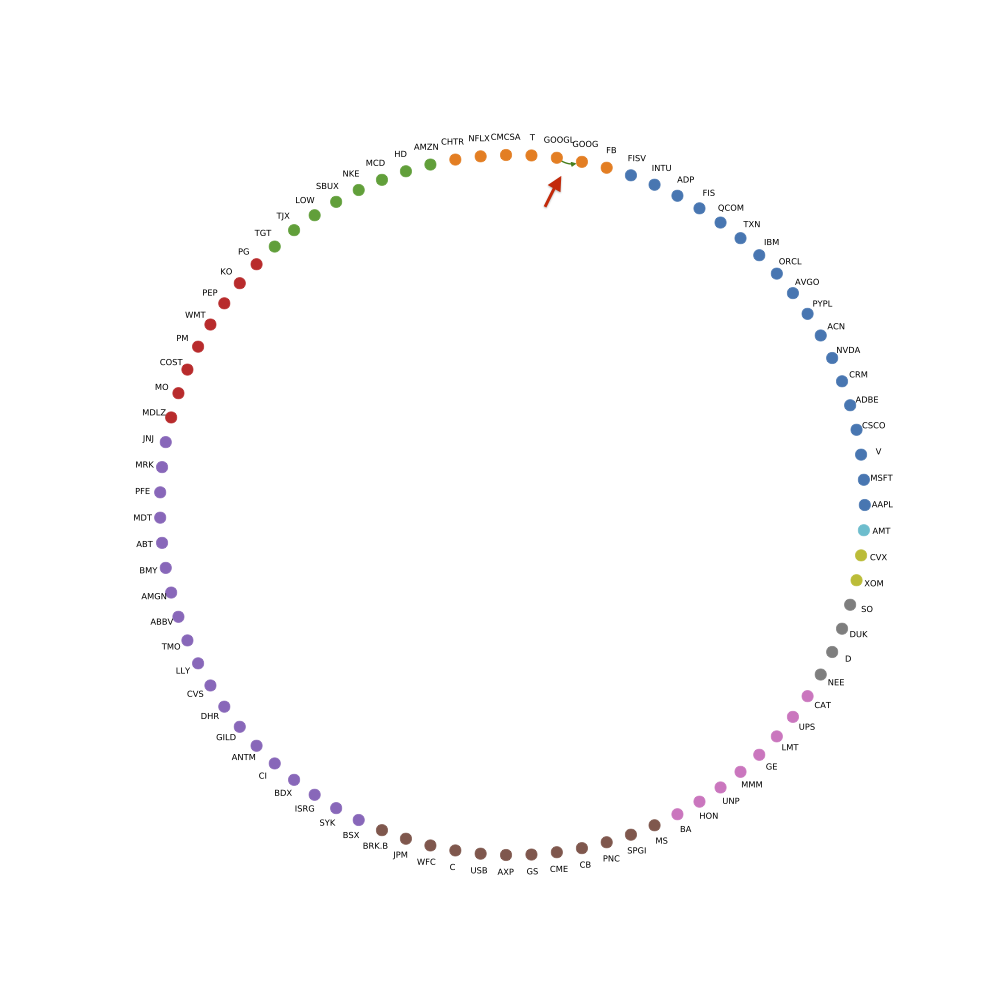}
}
\subfigure[Threshold = 75-th percentile, based on integrated OFIs  \label{fig:freq_lasso75_sec_int}]{
\includegraphics[width=.445\textwidth, trim=2.6cm 3cm 2.6cm 3cm,clip]{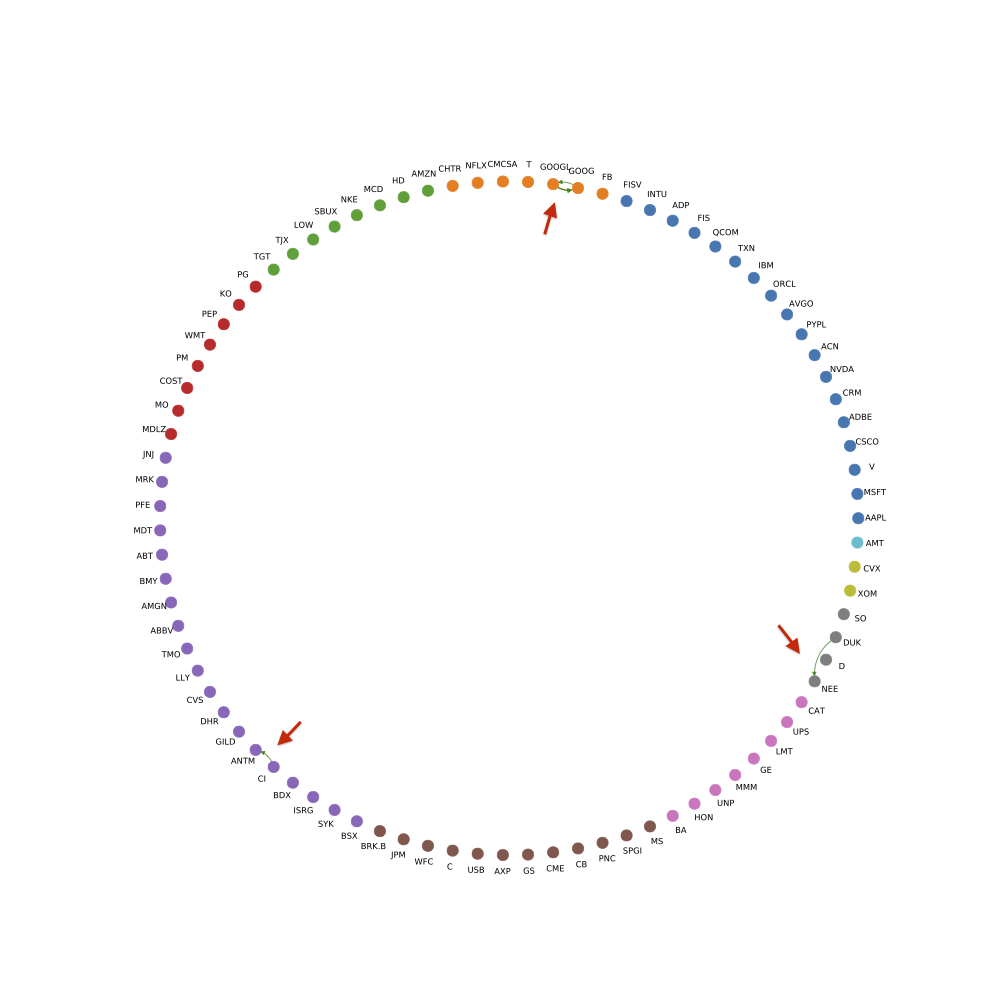}
}
\subfigure[Threshold = 25-th percentile, based on integrated OFIs  \label{fig:freq_lasso25_sec_int}]{
\includegraphics[width=.45\textwidth, trim=3.6cm 4cm 3.6cm 4cm,clip]{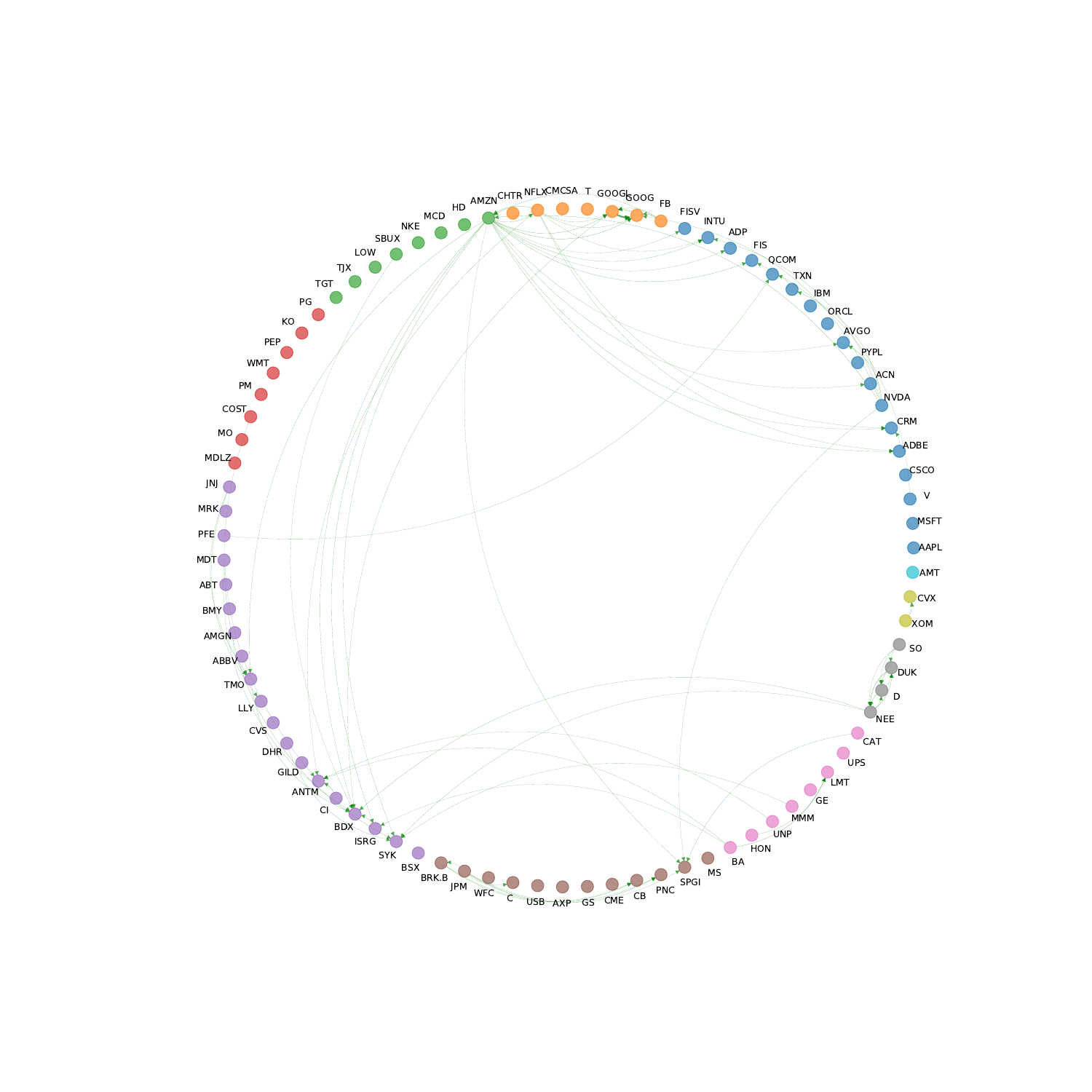}
}
\caption{Illustrations of the coefficient networks constructed from {contemporaneous cross-impact} models. }
\caption*{\textit{Note:} To render the networks more interpretable and for ease of visualization, we only plot the top 5\% largest (a-b), or top 25\% largest (c), or top 75\% largest (d), in magnitude coefficients. The coefficients are averaged over each regression window between 2017–2019. Nodes are colored by the GICS structure and sorted by market capitalization. Green links represent positive values while black links represent negative values. The width of edges is proportional to the absolute values of their respective coefficients.}
\label{fig:coef_lasso}
\end{figure}

We visualize a network for each coefficient matrix, which only preserves the edges larger than a given threshold (following \citet{kenett2010dominating, curme2015emergence}), as shown in Figure \ref{fig:coef_lasso}. We color stocks according to the GICS sector division, and sort them by their market capitalization within each sector.\footnote{The Global Industry Classification Standard (GICS) is an industry taxonomy developed in 1999 by MSCI and Standard \& Poor's (S\&P) for use by the global financial community.} As one can see from Figure \ref{fig:freq_lasso_sec}, the cross-impact coefficient matrix $\left[\beta_{i, j}^{[1]} \right]_{j \neq i}$ displays a sectorial structure, in accordance with previous studies (e.g. \citet{benzaquen2017dissecting}). This behavior could be fueled by index arbitrage strategies, where traders may, for example, trade an entire basket of stocks coming from the same sector against an index.

Figure \ref{fig:freq_lasso_sec_int} presents the network of cross-impact coefficients based on integrated OFIs, i.e. $\left[\beta_{i, j}^{I} \right]_{j \neq i}$. Compared with Figure \ref{fig:freq_lasso_sec}, the connections in Figure \ref{fig:freq_lasso_sec_int} are much weaker, implying that the cross-impact from stocks can be potentially explained by a stock's own multi-level OFIs, to a large extent.
Note that there is only one connection from GOOGL to GOOG, as pointed out at the top of Figure \ref{fig:freq_lasso_sec_int}. This stems from the fact that both stock ticker symbols pertain to Alphabet (Google). Our study also reveals that OFIs of GOOGL have more influence on the returns of GOOG, not the other way around. The main reason might be that GOOGL shares have voting rights, while GOOG shares do not.

In Figures \ref{fig:freq_lasso75_sec_int} and \ref{fig:freq_lasso25_sec_int}, we set lower threshold values (75-th, respectively, 25-th percentile of coefficients) {in order to promote more edges} in the networks based on integrated OFIs. Interestingly, we observe only four connections in Figure \ref{fig:freq_lasso75_sec_int}. Except from bidirectional links between GOOGL and GOOG, there exists a one-way link from Cigna (CI) to Anthem (ANTM), and another one-way link from Duke Energy (DUK) to NextEra Energy (NEE). Anthem announced to acquire Cigna in 2015. After a prolonged breakup, this merger finally failed in 2020. Therefore, it is unsurprising that the OFIs of Cigna can affect the price movements of Anthem. Conversely, Anthem's OFIs also have an impact on the price movements of Cigna, but to a lesser extent. Further research should be undertaken to investigate this phenomenon. In terms of the second pair, Duke Energy rebuffed NextEra's acquisition interest in 2020. Note that 2020 is not in our sample period. This finding  hints that certain market participants may have noticed the special relationship between Duke Energy and NextEra Energy before this mega-merger was proposed.

\subsubsection{Out-of-sample performance}
Although the in-sample estimation yields interesting findings, practitioners are eventually concerned about the out-of-sample estimation. 
Therefore, we propose to perform the following out-of-sample tests. We use the above fitted models to estimate returns on the following 30-minute data and compute the corresponding $R^2$, denoted as \textbf{out-of-sample $R^2$} or \textbf{OS $R^2$}.\footnote{Previous studies either investigated the in-sample $R^2$ (including \citet{cont2014price, capponi2020multi}), or adopted a cross-validation method (e.g. \citet{xu2018multi}). However, these works failed to consider the generalization error of their models or damaged the chronological order of the time-series data. In contrast, we obey the temporal ordering in our study. These matters are vital to practitioners, as only the historical data are accessible for the model fit in practice.}

Table \ref{tab:pi_ci_oos} reports the average values and their standard deviations of {out-of-sample $R^2$} of $\textbf{PI}^{[1]}$, $\textbf{CI}^{[1]}$, $\textbf{PI}^{I}$, and $\textbf{CI}^{I}$. We first focus on the models using best-level OFIs. It appears $\textbf{CI}^{[1]}$ has a slight advantage compared with $\textbf{PI}^{[1]}$ for out-of-sample tests with an improvement of 1.39\% (=66.03\%-64.64\%). However, when involving multi-level or integrated OFIs, the performance of $\textbf{CI}^{I}$ is slightly worse than $\textbf{PI}^{I}$, indicating that the cross-impact model with integrated OFIs cannot provide extra explanatory power to the price impact model with integrated OFIs. Overall, we observe that the models using integrated OFIs unveil significant and consistent improvements over those using only best-level OFIs.  

\begin{table}[H]
    \centering
    \caption{{Out-of-sample} performance for contemporaneous returns.}
    \resizebox{0.5\textwidth}{!}{

\begin{tabular}{lcccc}
    \toprule
    & \multicolumn{2}{c}{{Best-level OFIs}} & \multicolumn{2}{c}{{Integrated OFIs}} \\
     \cmidrule(lr){2-3}\cmidrule(lr){4-5} 
 & $\textbf{PI}^{[1]}$ &  $\textbf{CI}^{[1]}$ & $\textbf{PI}^{I}$ & $\textbf{CI}^{I}$ \\\midrule
OS $R^2$ & 64.64  & 66.03 & \text{83.83} & 83.62 \\
& (21.82) & (19.51) & (16.90) & (14.53) \\
\bottomrule
\end{tabular}}
    \caption*{\textit{Note:} The table reports the mean values and standard deviations (in parentheses) of {out-of-sample} $R^2$ (in percentage points) of various models when modeling contemporaneous returns. The models include  $\textbf{PI}^{[1]}$ (Eqn \eqref{eq:pi_best}), $\textbf{CI}^{[1]}$ (Eqn \eqref{eq:ci_best}), $\textbf{PI}^{I}$ (Eqn \eqref{eq:pi_int}), and $\textbf{CI}^{I}$ (Eqn \eqref{eq:ci_int}). These statistics are averaged across each stock and each regression window.}
    \label{tab:pi_ci_oos}
\end{table}

In general, we observe strong evidence implying $\textbf{CI}^{[1]}$ provides a better out-of-sample estimate than $\textbf{PI}^{[1]}$, while for $\textbf{PI}^{I}$ and $\textbf{CI}^{I}$, the evidence is opposite. However, it is important to note that these conclusions are based on a point estimate and do not necessarily indicate statistical significance. 
Therefore, we perform the following hypothesis test for each stock on the out-of-sample data to assess statistical significance,
\begin{equation*}
\mathcal{H}_{0}: \mathbb{E}\left[R^2_\mathrm{OS}\left(\textbf{CI}^{[1]}\right)-R^2_\mathrm{OS}\left(\textbf{PI}^{[1]}\right)\right] \leq 0 \text{  vs.  } \mathcal{H}_{1}: \mathbb{E}\left[R^2_\mathrm{OS}\left(\textbf{CI}^{[1]}\right)-R^2_\mathrm{OS}\left(\textbf{PI}^{[1]}\right)\right]>0.
\end{equation*}

We employ the approach from \citet{giacomini2006tests} and \citet{chinco2019sparse} to assess statistical significance through a Wald-type test (see \citet{ward2018maximum}). Theorem 1 in \citet{giacomini2006tests} implies that we can use a standard $t$-test to evaluate the statistical significance of changes in $R^2$.  A $p$-value less than a given significance level $\alpha$ rejects the null hypothesis in favor of the alternative at the $1-\alpha$ confidence level, implying $\textbf{CI}^{[1]}$ has significantly better estimation than $\textbf{PI}^{[1]}$. We also implement this test for the comparison between $\textbf{PI}^{I}$ and $\textbf{CI}^{I}$.

\begin{figure}[H]
    \centering
    \subfigure[$R^2_\mathrm{OS}\left(\textbf{CI}^{[1]}\right)-R^2_\mathrm{OS}\left(\textbf{PI}^{[1]}\right)$ \label{fig:pvalue_ci1}]{
    \includegraphics[width=0.96\textwidth, trim=4.5cm 2mm 4cm 2cm,clip]{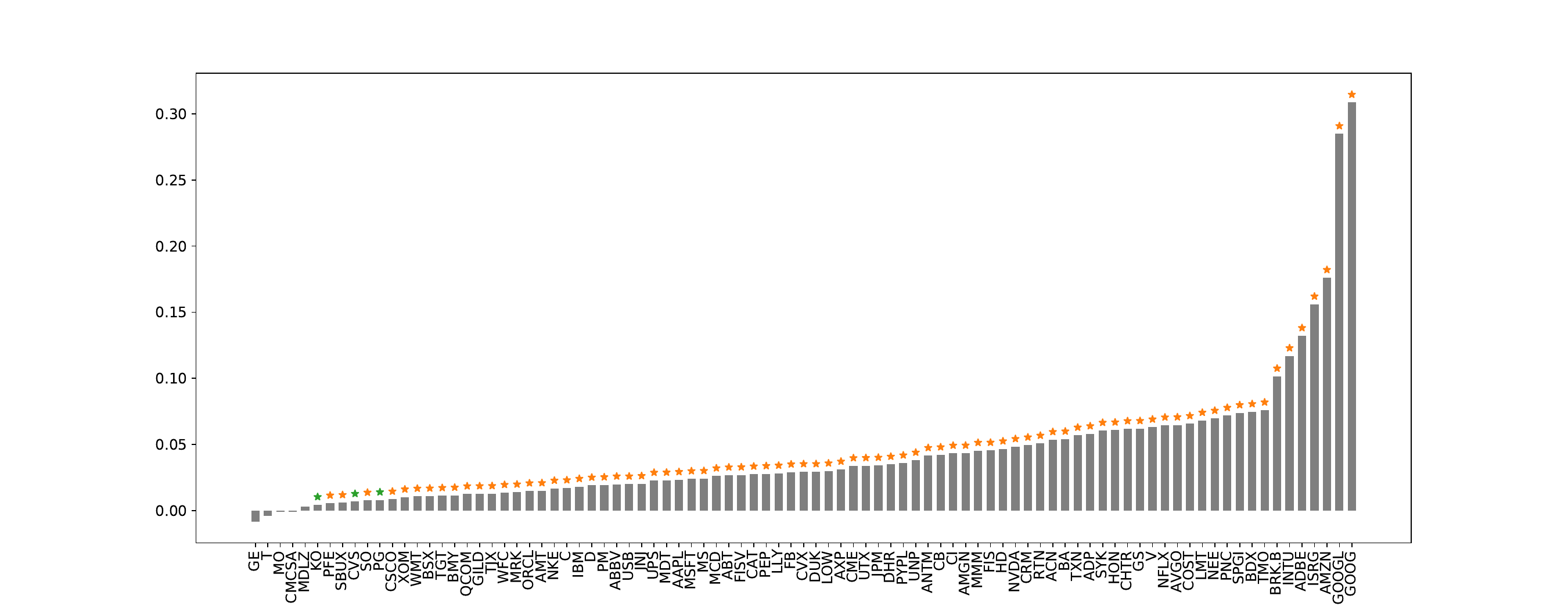}}
    \subfigure[$R^2_\mathrm{OS}\left(\textbf{CI}^{I}\right)-R^2_\mathrm{OS}\left(\textbf{PI}^{I}\right)$
    \label{fig:pvalue_ciI}]{
    \includegraphics[width=0.96\textwidth, trim=4.5cm 2mm 4cm 2cm,clip]{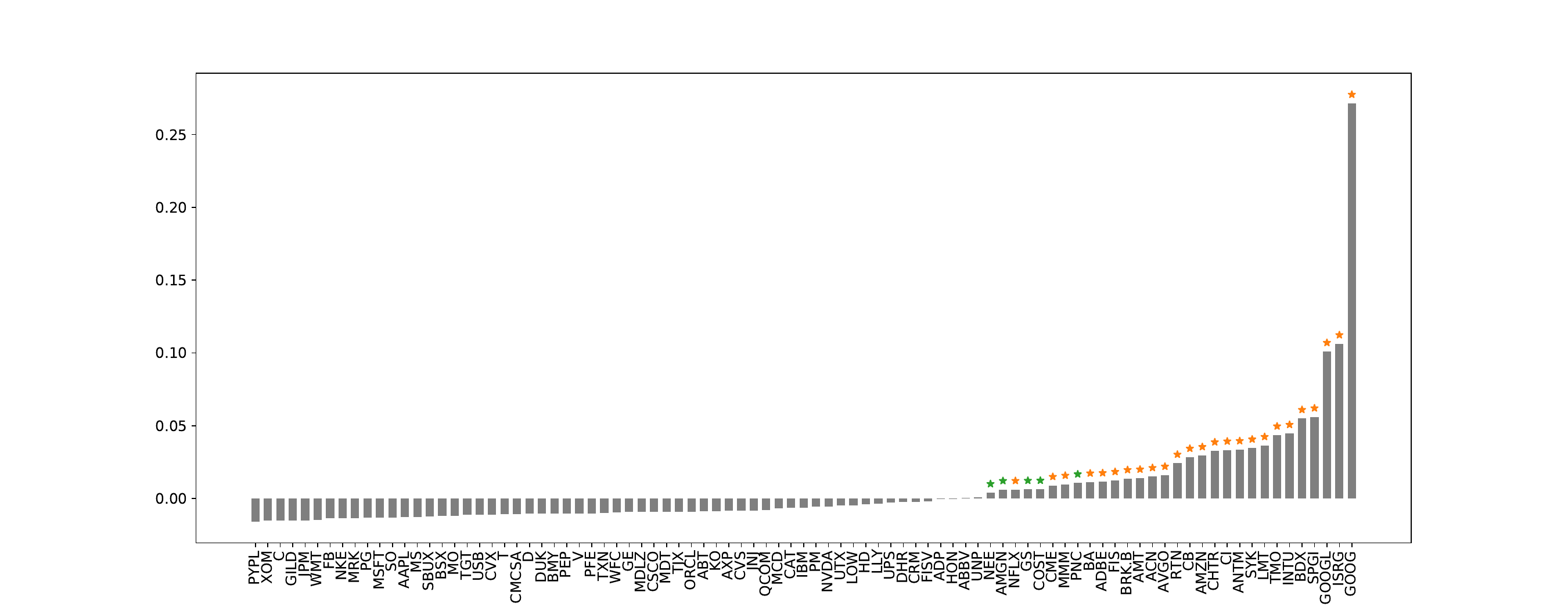}}
    \caption{Mean differences of out-of-sample $R^2$ between CI and PI models.}
    \caption*{\textit{Note:} A positive (negative) number indicates superiority for the CI (PI) model. The $y$-axis represents the average difference of OS $R^2$ between CI and PI, while the $x$-axis lists the stock symbols. Stars indicate the $p$-values, with orange, green, and blue representing significance at the 1\%, 5\%, and 10\% levels, respectively.}
    \label{fig:pvalue}
\end{figure}

Figure \ref{fig:pvalue} illustrates the main results from the above hypothesis tests. When using only the best-level OFIs, the cross-impact model is superior to the price impact model for 91.0\% (94.4\%) of stocks, at the 1\% (5\%) confidence level. However, when examining the models using integrated OFIs, we reject the null hypothesis (i.e., in favor of the cross-impact model) only for 28.1\% (33.7\%) of stocks at the 1\% (5\%) confidence level. As expected, cross-impact terms can significantly improve the explanatory power of the price impact model for GOOG and GOOGL.

{Dynamics of limit order book may depend on the tick-to-price ratio, or alternatively, the fraction of time that the bid-ask spread is equal to one tick for a given stock (\citet{curato2015tick}). We   examine whether this dependence also extends to cross-asset OFIs.\footnote{We would like to thank an anonymous reviewer for suggesting this analysis.} Our findings, presented in Table \ref{tab:r2_pi_ci_tick}, suggest that cross-asset OFIs can better explain the price dynamics of stocks with a larger tick-to-price ratio.}

\begin{table}[H]
    \centering
    \caption{Out-of-sample $R^2$ of various {contemporaneous} models sorted by tick-to-price ratio.} 
    \resizebox{0.6\textwidth}{!}{\begin{tabular}{ccccc}\toprule
  &  $[0\%, 25\%)$ & $[25\%, 50\%)$  &  $[50\%, 75\%)$ & $[75\%, 100\%]$ \\\midrule
$\textbf{PI}^{[1]}$ & 44.38 & 62.51 & 77.55 & 70.70 \\
$\textbf{CI}^{[1]}$ & 53.01 & 66.32 & 72.50 & 78.34 \\\midrule
$\textbf{PI}^{I}$   & 68.14 & 84.58 & 88.14 & 89.86 \\
$\textbf{CI}^{I}$   & 72.01 & 84.80 & 88.51 & 91.01 \\
\bottomrule
\end{tabular}}
    \caption*{\textit{Note:} $[0\%, 25\%)$, respectively $[75\%, 100\%]$, denote the subset of stocks with the lowest, respectively highest, 25\% values according to the tick-to-price ratio.}
    \label{tab:r2_pi_ci_tick}
\end{table}

\subsection{Discussion about contemporaneous cross-impact}

\subsubsection{Impact on stocks}
{In summary, our previous results mainly show that when considering only the best-level OFI of a single stock, the addition of the best-level OFI from other stocks slightly increases the explanatory power. On the other hand, when the information from multiple levels is integrated into the OFI, the improvement is negligible. In the meantime, it is unsurprising that taking into account more levels in the LOB ($\textbf{PI}^{I}$) could better explain price changes, compared to only considering best-level orders ($\textbf{PI}^{[1]}$).} 

{After observing these results, several natural questions may arise: How can the above facts be reconciled?\footnote{One possible explanation for those facts is that the duration of the cross-impact terms might be shorter than the current time interval (30 minutes) used in our experiments, rendering the cross-impact terms vanish in out-of-sample tests. To verify this  assertion, we implement additional experiments where models are updated more frequently. The results (deferred to Appendix \ref{sec_ci:ci_sup}) reveal that even under higher-frequency updates (1-min) of the models, there is no benefit from introducing cross-impact terms to the price impact model with integrated OFIs.} How do the cross-asset best-level OFIs interact with the multi-level OFIs, when modeling contemporaneous returns?}

\begin{figure}[H]
\centering
\includegraphics[width=0.95\textwidth]{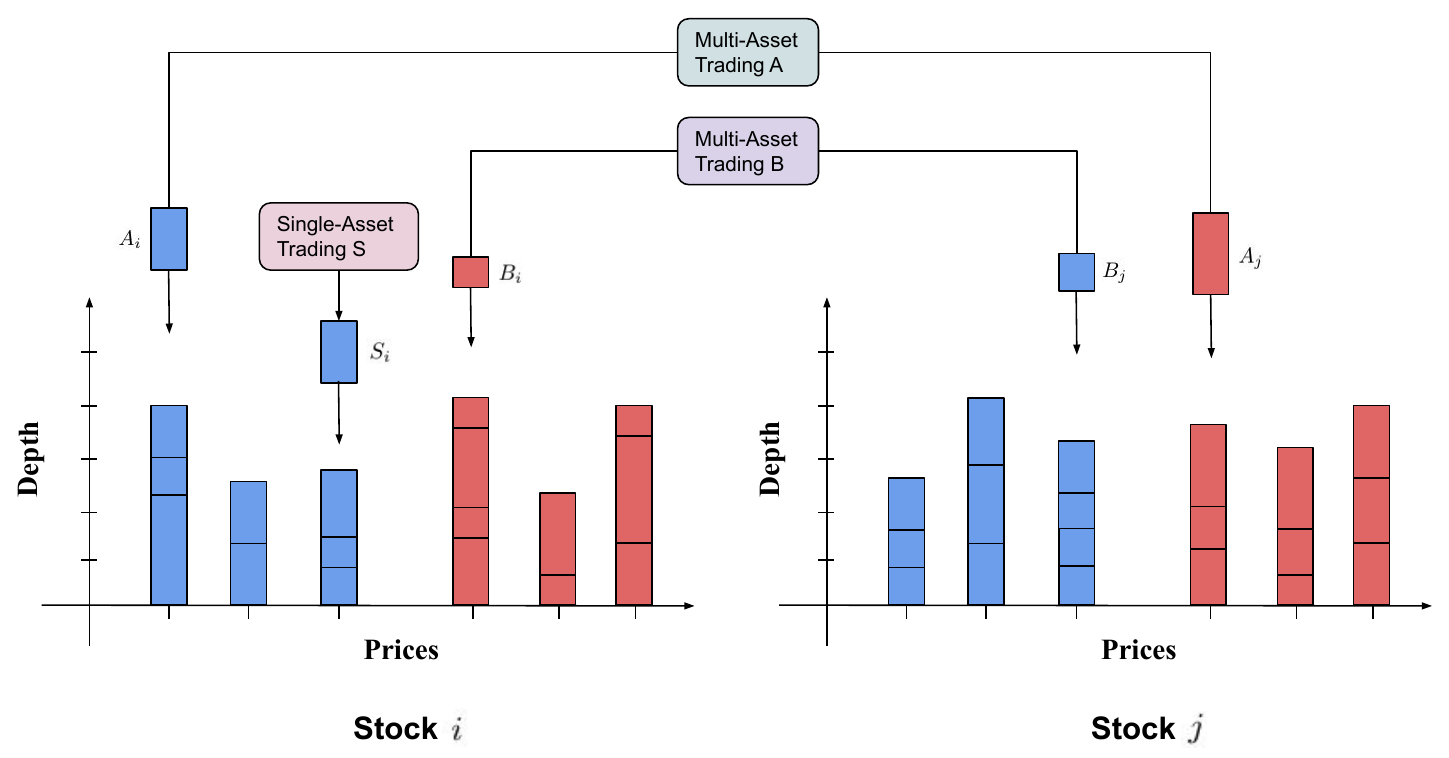}
\caption{Illustration of the cross-impact model.}
\caption*{\textit{Note:} The orders at different levels of each stock may come from single-asset and multi-asset trading strategies. The returns of stock $i$ are potentially influenced by orders of stock $j$ through the connections $B_j \to B_i \to \text{ofi}_i^{1} \to r_i$ and $A_j \to A_i \to \text{ofi}_i^{3} \to r_i$. Information along the path $A_j \to A_i \to \text{ofi}_i^{3} \to r_i$ can be collected by the price impact model with integrated OFIs but not by the price impact model with only best-level OFIs. }
\label{fig:CI_LOB}
\end{figure}

To address these questions, we consider the following scenario, also depicted in Figure \ref{fig:CI_LOB}. For simplicity, we denote the order from trading strategy $A$ on stock $i$ (respectively, $j$) as $A_i$ (respectively, $A_j$). Analogously, we define orders from strategy $B$ and $S$. Let us next consider the orders of stock $i$. There are three orders from different portfolios, given by $A_i$,  $B_i$ and $S_i$. $A_i$ is at the third bid level of stock $i$ and linked to an order at the best ask level of stock $j$, i.e. $A_j$. Also, $B_i$ is at the best ask level of stock $i$ and linked to an order at the best bid level of stock $j$, i.e. $B_j$. Finally, $S_i$ is an individual bid order at the best level of stock $i$.  

We observe that the best-level limit orders from stock $j$ may be linked to price movements of stock $i$ through paths $B_j \to B_i \to \text{ofi}_i^{1} \to r_i$ and $A_j \to A_i \to \text{ofi}_i^{3} \to r_i$. Thus the price impact model for stock $i$ which only utilizes its own best-level orders will ignore the information of $A_i$, while the cross-impact model can partially collect it along the path $A_j \to A_i$.  This might illustrate why the best-level OFIs of multiple assets can provide slightly additional explanatory power to the price impact model using only the best-level OFIs.

Nonetheless, if we can integrate multi-level OFIs in an efficient way (in our example, aggregate order imbalances caused by orders $A_i$, $B_i$  and $S_i$), then there is no need to consider OFIs from other stocks for modeling price dynamics. For example, information hidden in the  path $A_j \to A_i \to \text{ofi}_i^{3}  \to r_i$ can be leveraged as long as $A_i$ is well absorbed into new integrated OFIs. {In this sense, for stock $i$, cross-asset best-level OFIs (including $A_j$) are surrogates of its own OFIs at different levels (here $A_i$), to a certain extent. The likelihood of this relationship is attributed to massive portfolio trades that submit or cancel limit orders across a variety of assets at different levels.\footnote{\citet{cao2009information, hautsch2012market, sirignano2019deep, chakrabarty2022order} showed that the depth of some deeper levels (such 2-3) is higher than the best level depth.}} We put forward this mechanism which potentially explains why the cross-impact model with integrated OFIs cannot provide additional explanatory power compared to the price impact model with integrated OFIs.\footnote{It would be a very interesting research direction to derive testable predictions for this mechanism in future work. However, so far it hinges on the availability of client-ID based LOB data, i.e. knowing that the same market participant is behind the orders for two related instruments, released at approximately the same time, such as $(A_i, A_j)$, $(B_i, B_j)$.}

\subsubsection{Impact on portfolios}

{A related question is about the aggregation of cross-impact at portfolio level. Let us consider the OFI of portfolio $p$ as $\mathrm{ofi}_{p, t}^{1, h} := \sum_{i=1}^N w_i \mathrm{ofi}_{i, t}^{1, h}$, where $w_i$ is the weight of asset $i$ in a portfolio. Then the price impact for portfolio $p$ is 
\begin{align}\label{eq:pi_portfolio}
    r_{p, t}^{(h)} &= a_{p}^{[1]} + \beta_{p}^{[1]} \mathrm{ofi}_{p, t}^{1, h}+e_{p, t}^{[1]} \\\nonumber
    &= a_{p}^{[1]} + \beta_{p}^{[1]} \sum_{i=1}^N w_i \mathrm{ofi}_{i, t}^{1, h}+e_{p, t}^{[1]},
\end{align}
where $a_{p}^{[1]}$ is the intercept and $e_{p, t}^{[1]}$ is the noise term.}

{On the other hand, the cross-impact for portfolio $p$ is 
\begin{align}\label{eq:ci_portfolio}
r_{p, t}^{(h)} &= \sum_{i=1}^N w_i r_{p, t}^{(h)} \\\nonumber
&= \sum_{i=1}^N w_i \left( \alpha_{i}^{[1]} + \beta_{i}^{[1]} \mathrm{ofi}_{i, t}^{1, h}+\epsilon_{i, t}^{[1]} \right) \\\nonumber
&= \alpha_{p}^{[1]} + \sum_{i=1}^N \beta_{i}^{[1]} w_i  \mathrm{ofi}_{i, t}^{1, h} + \epsilon_{p, t}^{[1]},
\end{align}
where $\alpha_{p}^{[1]} = \sum_{i=1}^N w_i \alpha_{i}^{[1]}$ is the intercept.}

{Comparing  \eqref{eq:pi_portfolio} and \eqref{eq:ci_portfolio} shows that  cross-impact at  portfolio level depends on the angles of $\vec{\beta}=(\beta_{1}^{[1]},\dots,\beta_{N}^{[1]})$ and  $\vec{w}=(w_1, \dots, w_N)$. On one hand, if individual assets exhibit a universal pattern of price dynamics, i.e. $\beta_{i}^{[1]} \approx \beta_{j}^{[1]}, \forall i \neq j$, we may conclude that the portfolio return is driven by its own order flows, to a large extent. On the other hand, if the portfolio places most of the weight on a specific stock or a set of stocks with similar patterns of price dynamics, then the portfolio returns are also driven by its own order flows, rather than by cross-impact. In other scenarios, it is necessary to consider the cross-impact. One can perform an analogous analysis for models with integrated OFIs.}

{In fact, the mechanism depicted in Figure \ref{fig:CI_LOB} aligns with this analysis. For example, orders placed on stock $i$ represented by $A_i$ and $S_i$ have the potential to affect the price movements of that stock, which subsequently affect the returns of portfolio $B$, even if $A_i$ and $S_i$ have no direct association with $B$. Additionally, $A_j$ may have a different influence on the performance of $B$ because of the (potentially) different price dynamics of stock $j$. Therefore, it may be necessary to take into account cross-asset OFIs when developing models for portfolio returns.} 

{To examine the potential of cross-asset OFIs in explaining portfolio returns, we choose two widely-used portfolio construction methods: the equal-weighted portfolio (EW), and the eigenportfolio\footnote{See \citet{avellaneda2010statistical}.} using the 1st principal component (Eig1). Table \ref{tab:port_pi_ci} summarizes the out-of-sample $R^2$ of various models on these two portfolios. As Table \ref{tab:port_pi_ci} shows, there is a significant difference between $\textbf{PI}^{[1]}$ and $\textbf{CI}^{[1]}$, $\textbf{PI}^{I}$ and $\textbf{CI}^{I}$, indicating that cross-impact is a crucial factor when modeling portfolio returns. Again, the models using the integrated OFIs outperform their counterparts using the best-level OFIs.}

\begin{table}[H]
    \centering
    \caption{Out-of-sample performance on portfolio returns.} 
    \resizebox{0.48\textwidth}{!}{\begin{tabular}{lcccc}
    \toprule
    & \multicolumn{2}{c}{{Best-level OFIs}} & \multicolumn{2}{c}{{Integrated OFIs}} \\
     \cmidrule(lr){2-3}\cmidrule(lr){4-5} 
 & $\textbf{PI}^{[1]}$ &  $\textbf{CI}^{[1]}$ & $\textbf{PI}^{I}$ & $\textbf{CI}^{I}$ \\\midrule
EW & 79.29  & 81.03 & 85.26 & \text{87.97} \\
& (6.24) & (6.16) & (3.54) & (2.78) \\\midrule
Eig1 & 80.73  & 81.95 & 84.69 & \text{87.70} \\
& (6.75) & (6.36) & (3.70) & (2.84) \\
\bottomrule
\end{tabular}}
    \caption*{\textit{Note:} EW: equal-weighted portfolio. Eig1: eigenportfolio using the 1st principal component of multi-asset returns.}
    \label{tab:port_pi_ci}
\end{table}


\section{Forecasting future returns} \label{sec_ci:forward}
In the previous section, the definitions of price impact and cross-impact are based on contemporaneous OFIs and returns, meaning that both quantities pertain to the same bucket of time. In this section, we extend the above studies to future returns, and probe into the forward-looking price impact and cross-impact models.

\subsection{Predictive models}
We first propose the following forward-looking price impact and cross-impact models, denoted as $\textbf{FPI}^{[1]}$, $\textbf{FPI}^{I}$, $\textbf{FCI}^{[1]}$, and $\textbf{FCI}^{I}$, respectively.  $\textbf{FPI}^{[1]}$ ($\textbf{FPI}^{I}$) uses the lagged best-level (integrated) OFIs of stock $i$ to predict its own future return $r_{i, t+f}^{(f)}$ during $(t, t+f]$, while $\textbf{FCI}^{[1]}$ ($\textbf{FCI}^{I}$) involves the lagged multi-asset best-level (integrated) OFIs. We employ OLS to fit the forward-looking price impact models and LASSO to fit the cross-impact models.
\begin{align}
\textbf{FPI}^{[1]}:\quad r_{i, t+f}^{(f)} &= \alpha_{i}^{[1]} + \sum_{k \in L} \beta_{i}^{[1], k} \mathrm{ofi}_{i, t}^{1, (kh)} +\epsilon_{i, t+f}^{[1]}, \label{eq:for_pi}\\
\textbf{FCI}^{[1]}:\quad r_{i, t+f}^{(f)} &= \alpha_{i}^{[1]} + \sum_{j=1}^{N} \sum_{k \in L}  \beta_{i,j}^{[1], k} \mathrm{ofi}_{i, t}^{1, (kh)}  + \eta_{i, t+f}^{[1]} \label{eq:for_ci}, \\
\textbf{FPI}^{I}:\quad r_{i, t+f}^{(f)} &= \alpha_{i}^{I} + \sum_{k \in L} \beta_{i}^{I, k} \mathrm{ofi}_{i, t}^{I, (kh)} +\epsilon_{i, t+f}^{I}, \label{eq:for_pi_int}\\
\textbf{FCI}^{I}:\quad r_{i, t+f}^{(f)} &= \alpha_{i}^{I} + \sum_{j=1}^{N} \sum_{k \in L}  \beta_{i,j}^{I, k} \mathrm{ofi}_{i, t}^{I, (kh)}  + \eta_{i, t+f}^{I} \label{eq:for_ci_int},
\end{align}
where $f$ is the forecasting horizon of future returns and $L=\{1, 2, 3, 5, 10, 20, 30\}$ represents the set of lags. 

Furthermore, we compare OFI-based models with return-based models studied in previous works, where lagged returns are involved as predictors. $\textbf{AR}$ (Eqn \eqref{eq:for_ar}) is an autoregressive (AR) model using various returns over different time horizons, 
inspired by \citet{corsi2009simple, ait2022and}. $\textbf{CAR}$ (\citet{chinco2019sparse}) uses the entire cross-section lagged returns as candidate predictors, as detailed in Eqn \eqref{eq:for_cr}. We employ OLS to fit the \textbf{AR}s and LASSO to fit the \textbf{CAR}s.
\begin{align}
\textbf{AR}:\quad r_{i, t+f}^{(f)} &= \alpha_{i} +  \sum_{k \in L} \beta_{i}^{r, k} r_{i, t}^{(kh)} +\epsilon_{i, t+f} \label{eq:for_ar} \\
\textbf{CAR}:\quad r_{i, t+f}^{(f)} &= \alpha_{i} + \sum_{j=1}^{N} \sum_{k \in L} \beta_{i,j}^{r, k} r_{i, t}^{(kh)} + \eta_{i, t+f} \label{eq:for_cr}
\end{align}
 
\subsection{Empirical results}
In this experiment, observations associated with returns and OFIs are computed minutely, i.e. $h=1$ minute.\footnote{{Note that we choose to use the physical time as opposed to the trading time. This is because each stock has its own specific trading time, which is asynchronous with that of others. Thus it is difficult to work out the cross–impact between stocks on a trading time scale, also see \citet{wang2016average, wang2016cross}. The choice of a 1-minute bin size allows us to abstract away from microstructure effects which are not the focus of the present mesoscopic study, as is the case in   \citet{benzaquen2017dissecting, chinco2019sparse}.}} Following \citet{chinco2019sparse}, we use data from the previous 30 minutes to estimate the model parameters and apply the fitted model to forecast future $f$-minute returns. We then move one minute forward and repeat this procedure to compute the rolling $f$-minute-ahead return forecasts. For all models, we initially focus on the 1-minute forecasting horizon. In Section \ref{sec_ci:forecast_longer}, we  consider return forecasts over longer horizons, including $f \in \{2, 3, 5, 10, 20, 30\}$ minutes, to assess the strength and duration of price impact and cross-impact.

Following the analysis of \citet{cartea2018enhancing, chinco2019sparse}, we demonstrate the effectiveness of the forward-looking price impact and cross-impact models from two perspectives: (1) statistical performance, and (2) economic gain. 

\subsubsection{Statistical performance}

Table \ref{tab:pi_ci_for} summarizes the out-of-sample $R^2$ values of the aforementioned predictive models when predicting the subsequent 1-minute returns, i.e. $f=1$. It appears the cross-impact models $\textbf{FCI}^{[1]}$ (respectively, $\textbf{FCI}^{I}$, \textbf{CAR}) achieve higher out-of-sample $R^2$ statistics compared to the price impact models $\textbf{FPI}^{[1]}$ (respectively, $\textbf{FPI}^{I}$, \textbf{AR}). We also implement the same hypothesis test described in Section \ref{sec_ci:ci} to investigate the statistical significance (unreported) of these results. We observe that the cross-impact models exhibit significantly superior performance than the price impact models across all stocks, at the 1\% confidence level.

\begin{table}[H]
\centering
\caption{Out-of-sample performance for one-minute-ahead returns.}
\resizebox{0.66\textwidth}{!}{

\begin{tabular}{lcccccc}
    \toprule
    & \multicolumn{2}{c}{{Best-level OFIs}} & \multicolumn{2}{c}{{Integrated OFIs}} & \multicolumn{2}{c}{{Returns}}\\
   \cmidrule(lr){2-3}\cmidrule(lr){4-5}\cmidrule(lr){6-7}
     
 & $\textbf{FPI}^{[1]}$ & $\textbf{FCI}^{[1]}$ & $\textbf{FPI}^{I}$ & $\textbf{FCI}^{I}$ & $\textbf{AR}$ & $\textbf{CAR}$ \\\midrule
OS $R^2$ & -0.37  & -0.10 & -0.36 & -0.10 & -0.36 & -0.10 \\
& (0.10) & (0.05) & (0.08) & (0.05) & (0.11) & (0.05) \\
\bottomrule
\end{tabular}
}
\caption*{\textit{Note:} The table reports the mean values and standard deviations (in parentheses) of {out-of-sample} $R^2$ of various models when modeling one-minute-ahead returns. The predictive models include $\textbf{FPI}^{[1]}$ (Eqn \eqref{eq:for_pi}), $\textbf{FCI}^{[1]}$ (Eqn \eqref{eq:for_ci}), $\textbf{FPI}^{I}$ (Eqn \eqref{eq:for_pi_int}), $\textbf{FCI}^{I}$ (Eqn \eqref{eq:for_ci_int}), \textbf{AR} (Eqn \eqref{eq:for_ar}) and \textbf{CAR} (Eqn \eqref{eq:for_cr}). These statistics are averaged across each stock and each regression window.}
\label{tab:pi_ci_for}
\end{table}

{Most of the empirical literature in return prediction focuses its evaluations on out-of-sample $R^2$. However, we remark that negative $R^2$ values do not imply that the forecasts are economically meaningless (see more discussions in \citet{kelly2022virtue, choi2022alpha}).\footnote{{A simple example can be framed as follows. Consider a model with one predictor and suppose that the estimated predictive coefficient is a significantly large multiple of the actual value. In this case, the $R^2$ will become negative. However, the predictions will be perfectly correlated with the true expected return, resulting in a positive expected return for our strategy. Proposition 4 in \citet{kelly2022virtue} further proposed that we can worry less about the positivity of out-of-sample $R^2$ from a prediction model and focus more on the out-of-sample performance of specific trading strategies based on predicted returns.}}} To emphasize this point, we will incorporate these return forecasts into a forecast-based trading strategy, and showcase their profitability in the following subsection.

Considering the different magnitudes of the OFIs and returns, we first normalize the coefficient matrix of each model by dividing by the average of the absolute coefficients. 
Figure \ref{fig:ci_for_coef_sec} (deferred to Appendix \ref{sec_ci:forward_add}) shows the average coefficient matrices of $\textbf{FCI}^{[1]}$, $\textbf{FCI}^{I}$, and \textbf{CAR}. For example, as revealed in Figure \ref{fig:ci_for_coef_sec}(a) ($\textbf{FCI}^{[1]}$), for a specific stock, the main influence comes from its own OFI, i.e. the absolute values of diagonal elements are significantly larger than the off-diagonal ones. We observe that cross-impact is often negative, consistent with \citet{pasquariello2015strategic}. Except for the self-impact, most stocks are also influenced by stocks in Communication Services, Consumer Discretionary and Information Technology. 

\begin{figure}[H]
    \centering
\subfigure[Based on best-level OFIs \label{fig:for_coef_lasso_sec_ofi1}]{
\includegraphics[width=.3\textwidth, trim=3.6cm 4cm 3.6cm 4cm,clip]{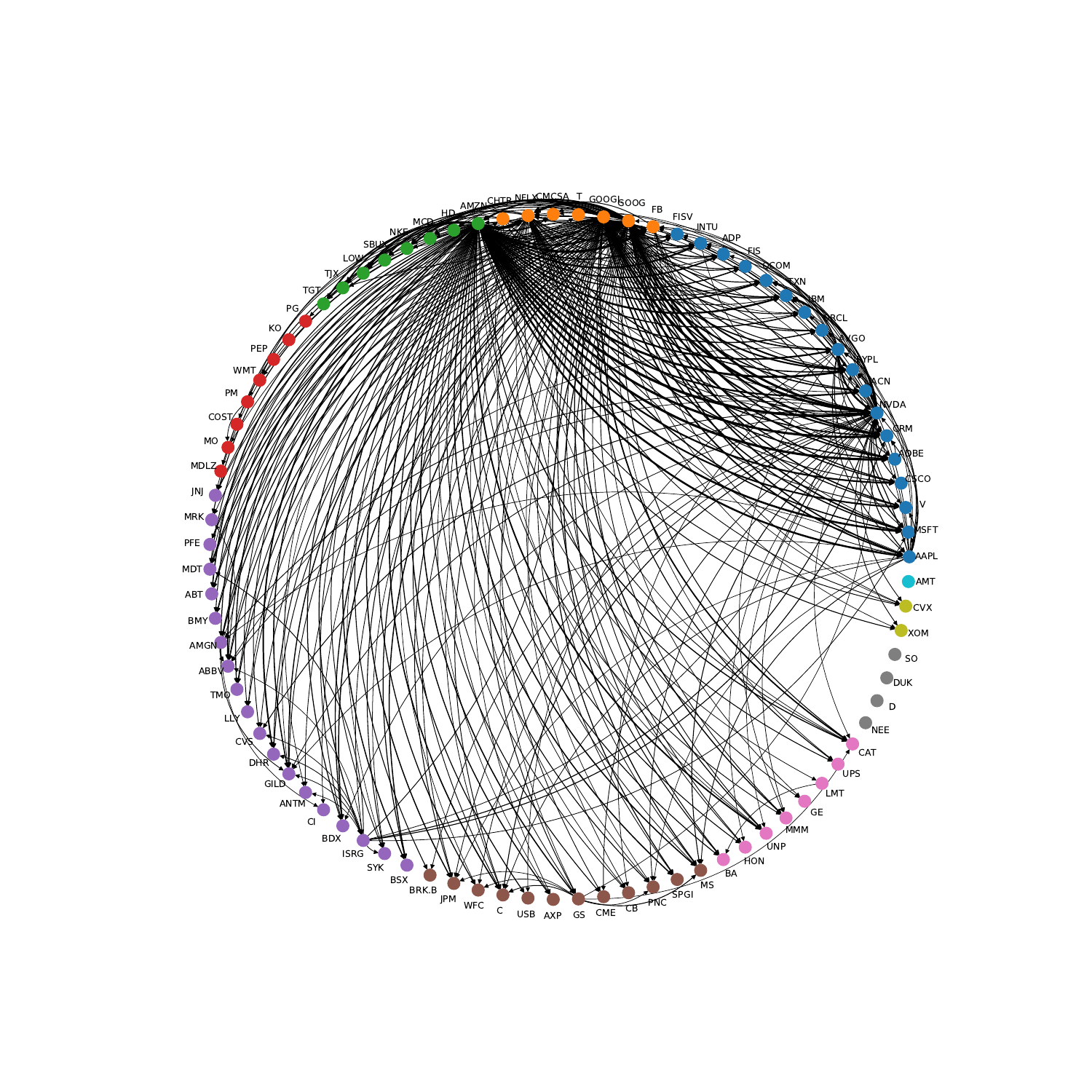}
}
\subfigure[Based on integrated OFIs\label{fig:for_coef_lasso_sec_ofiI}]{
\includegraphics[width=.3\textwidth, trim=3.6cm 4cm 3.6cm 4cm,clip]{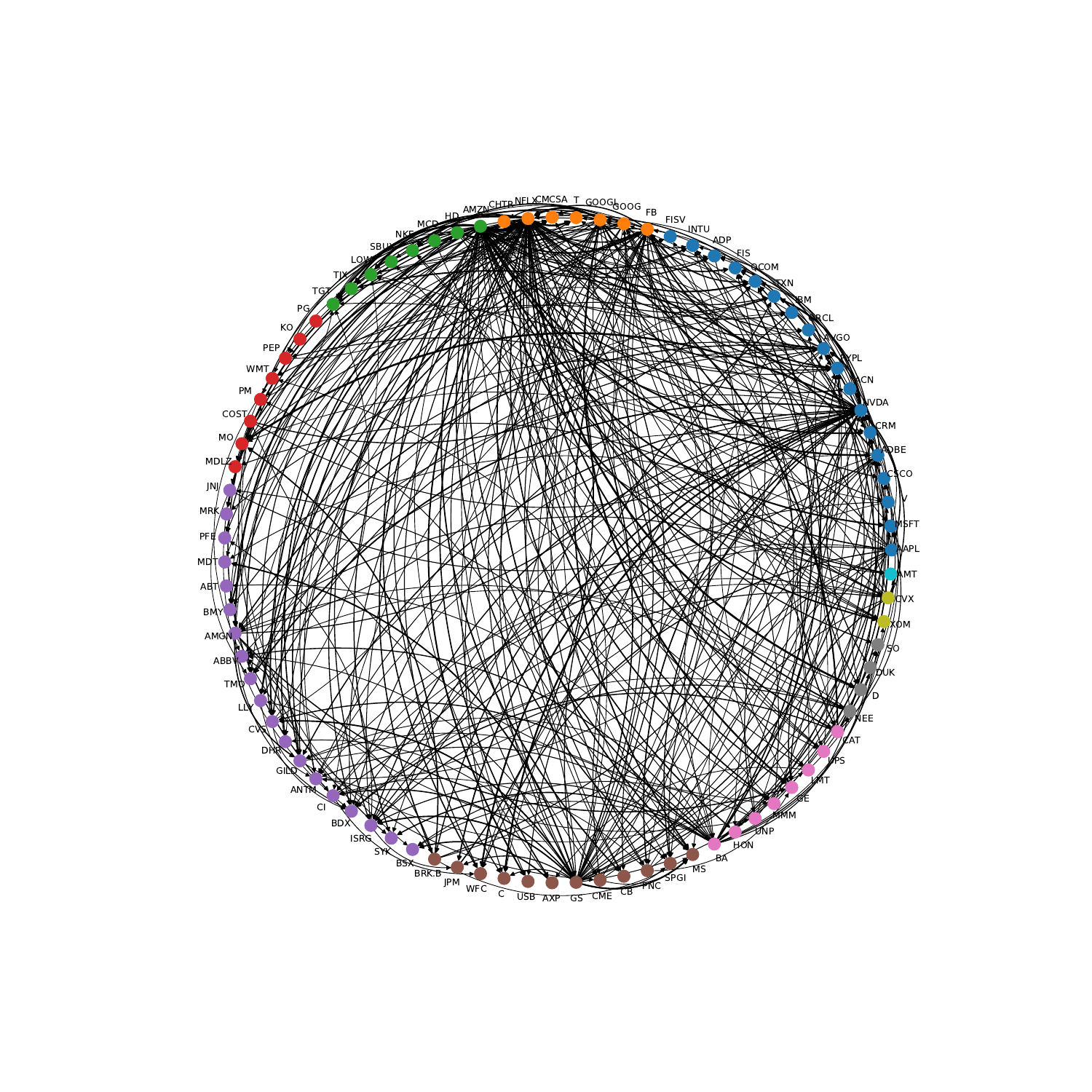}
}
\subfigure[Based on returns \label{fig:for_coef_lasso_sec_ret}]{
\includegraphics[width=.3\textwidth, trim=3.6cm 4cm 3.6cm 4cm,clip]{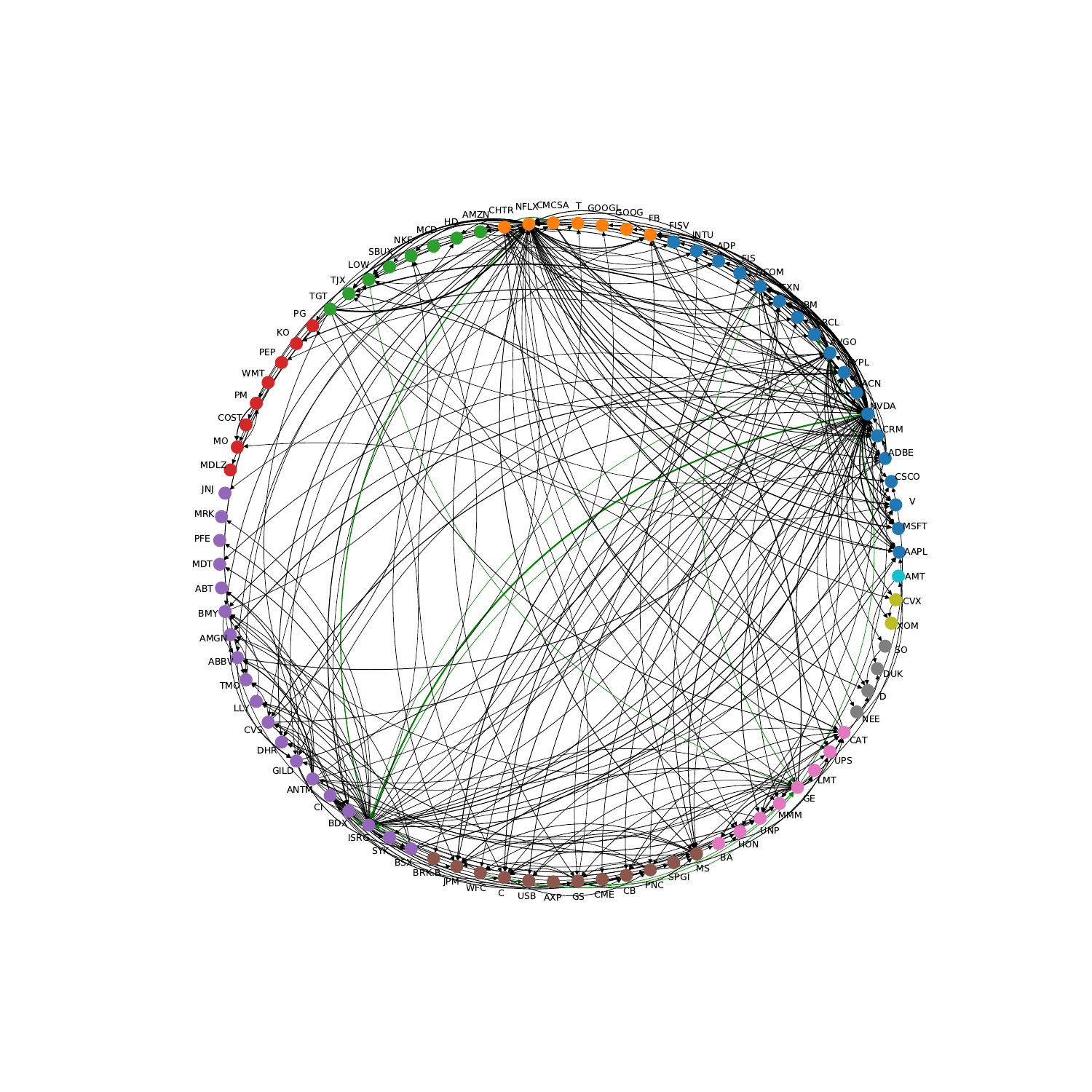}
}  
\caption{Network structure of the coefficient matrix constructed from {forward-looking cross-impact} models. }
\caption*{\textit{Note:} The coefficients are averaged over 2017–2019. To render the networks more interpretable and for ease of visualization, we only plot the top 5\% largest in magnitude coefficients. Nodes are colored by the GICS structure and sorted by market capitalization. Green links represent positive values while black links represent negative values. The width of edges is proportional to the absolute values of their respective coefficients.}
\label{fig:for_coef_lasso}
\end{figure}

To better illustrate the interactions between different stocks, we construct a network for each normalized coefficient matrix and only preserve the cross-asset edges (i.e. off-diagonal elements) larger than the 95-th percentile of coefficients. Figure \ref{fig:for_coef_lasso} illustrates some of the main characteristics of the coefficient networks for $\textbf{FCI}^{[1]}$, $\textbf{FCI}^{I}$, and \textbf{CAR}. For example, we again observe that there are more edges from Communication Services, Consumer Discretionary and Information Technology, indicating they may contain more predictive power for others.

\begin{table}[H]
\centering
\caption{Group degree centrality  for each GICS sector.}
\resizebox{1.0\textwidth}{!}{\begin{tabular}{lcccccc}\toprule
& \multicolumn{3}{c}{Group In-degree Centrality} & \multicolumn{3}{c}{Group Out-degree Centrality} \\\midrule
                       & Best-level OFIs          & Integrated OFIs    & Returns      & Best-level OFIs      & Integrated OFIs   & Returns         \\\midrule
\textcolor{colorA}{Information Technology}   & 0.12           & 0.36                        & 0.26               & 0.46           & 0.62                        & 0.59                    \\
\textcolor{colorB}{Communication Services} & 0.06           & 0.24                        & 0.20               & 0.85           & 0.74                        & 0.60                 \\
\textcolor{colorC}{Consumer Discretionary} & 0.09           & 0.20                        & 0.15               & 0.86           & 0.51                        & 0.17                \\
\textcolor{colorD}{Consumer Staples}       & 0.03           & 0.15                        & 0.09               & 0.00           & 0.11                        & 0.01                \\
\textcolor{colorE}{Health Care}            & 0.10           & 0.37                        & 0.19               & 0.12           & 0.22                        & 0.59                \\
\textcolor{colorF}{Financials}             & 0.12           & 0.21                        & 0.17               & 0.03           & 0.41                        & 0.08                \\
\textcolor{colorG}{Industrials}            & 0.10           & 0.19                        & 0.20                & 0.00           & 0.39                        & 0.27                \\
\textcolor{colorH}{Utilities}              & 0.00           & 0.07                        & 0.04               & 0.00           & 0.06                        & 0.00                \\
\textcolor{colorI}{Energy}                 & 0.06           & 0.07                        & 0.04               & 0.00           & 0.14                        & 0.00                \\
\textcolor{colorJ}{Real Estate}            & 0.00           & 0.05                        & 0.02               & 0.00           & 0.00                        & 0.01   \\\bottomrule   
\end{tabular}}
\caption*{\textit{Note:} According to the threshold networks as shown in Figure \ref{fig:for_coef_lasso}, we compute the fraction of stocks outside of a specific sector connected to stocks in this specific sector. The color of each sector in this table corresponds to the color in Figure \ref{fig:for_coef_lasso}.}
\label{tab:fut_group_degree}
\end{table}

To gain a better understanding of the structural properties of the resulting network, we aggregate node centrality measures (see \citet{everett1999centrality}) at the sector level, and also perform a spectral analysis of the adjacency matrix. From Table \ref{tab:fut_group_degree}, we observe that the out-degree centrality of Communication Services, Consumer Discretionary and Information Technology is significantly larger than that of others, consistent with previous findings. Figure \ref{fig:for_coef_lasso_sec_ret} also shows that the network based on returns contains more inner-sector connections than the other two counterparts, thus implying a sectorial structure. Table \ref{tab:fut_hubs} presents the top five most significant stocks in terms of out-degree centrality in each network, which exhibit more impact on the prices of other stocks.

\begin{table}[H]
\centering
\caption{Top 5 stocks according to node out-degree centrality in threshold networks. } 
\resizebox{.5\textwidth}{!}{\begin{tabular}{ccc}\toprule
Best-level OFIs          & Integrated OFIs    & Returns                \\\midrule
AMZN           & NFLX                           & NVDA                   \\
GOOG           & AMZN                           & NFLX                   \\
GOOGL          & NVDA                           & ISRG                   \\
NVDA           & GS                             & AVGO                   \\
NFLX           & FB                             & GE     \\\bottomrule 
\end{tabular}
}
\caption*{\textit{Note:} The out-degree centrality for a node is the fraction of nodes its outgoing edges are connected to.}
\label{tab:fut_hubs}
\end{table}

\begin{figure}[H]
    \centering
    \includegraphics[width=1.0\textwidth, trim=1cm 1mm 1cm 2mm,clip]{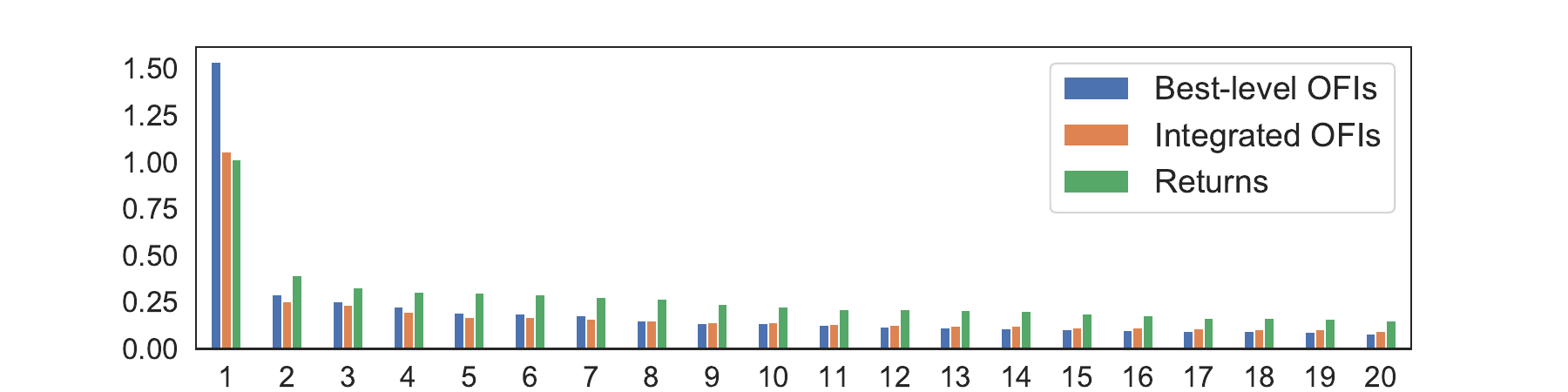}
    \caption{Barplot of normalized singular values for the average coefficient matrix in {forward-looking cross-impact} models.}
    \caption*{\textit{Note:} We perform Singular Value Decomposition (SVD) on the coefficient matrix and obtain the singular values. The $x$-axis represents the singular value rank,  and the $y$-axis shows  the normalized singular values. The coefficients are averaged over 2017–2019.}
    \label{fig:fut_singular}
\end{figure}

Figure \ref{fig:fut_singular} shows a barplot with the average value for the top 20 largest singular values of the network adjacency matrix, for best-level OFIs, integrated OFIs, and returns, where the average is performed over all constructed networks. For ease of visualization and comparison, we first normalize the adjacency matrix before computing the top singular values, which exhibit a fast decay.  
In addition to the significantly large top singular value revealing that the networks have a strong rank-1 structure, the next 6-8 singular values are likely to correspond to the more prominent industry sectors.

\subsubsection{Economic gains}
On the basis of return forecasts, we employ a portfolio construction method, proposed by \citet{chinco2019sparse}, to evaluate the economic gains of the aforementioned predictive models.

\paragraph{Forecast-implied portfolio.} For a specific forecasting model $F$, the motivations of portfolio construction can be summarized as follows.
\begin{itemize}
    \item It only executes an order when the one-minute-ahead return forecast exceeds the bid-ask spread.
    \item It buys/sells more shares of the $i$-th stock when the absolute value of one-minute-ahead return forecast for $i$-th stock is higher.
    \item It buys/sells more shares of the $i$-th stock when the one-minute-ahead return forecasts for the $i$-th stock tend to be less volatile throughout the trading day.
\end{itemize}

This strategy allocates a fraction $w_{i, t}$  of its capital to the $i$-th stock
\begin{equation}\label{eq:port_weight}
w_{i, t} \stackrel{\text { def }}{=} \frac{1_{\left\{\left|f_{i, t}^{F}\right|>s p r d_{i, t}\right\}} \cdot f_{i, t}^{F} \, / \, {\sigma}_{i, t}^{F}}{\sum_{n=1}^{N} 1_{\left\{\left|f_{n, t}^{F}\right|>s p r d_{n, t}\right\}} \cdot \left|f_{n, t}^{F}\right| \, / \, {\sigma}_{n, t}^{F}},
\end{equation}
where $f_{i, t}^{F}$ represents the one-minute-ahead return forecast according to model $F$ for minute $(t + 1)$, $sprd_{i, t}$ represents the relative bid-ask spread at time $t$, ${\sigma}_{i, t}^{F}$ represents the standard deviation of the model's one-minute-ahead return forecasts for the $i$-th stock during the previous 30 minutes of trading, i.e. the standard deviation of in-sample fits. The denominator is the total investment so that the strategy is self-financed. If there are no stocks with forecasts that exceed the spread in a given minute, then we set $w_{i, t}=0, \forall i$. 

Finally, we compute the \textit{profit and loss (PnL)} of the resulting portfolios on each trading day by summing the strategy's minutely returns as in \citet{chinco2019sparse}.

\begin{table}[H]
\centering
\caption{Economic performance of forecast-implied trading strategy.}
\resizebox{0.66\textwidth}{!}{

\begin{tabular}{lcccccc}
    \toprule
    & \multicolumn{2}{c}{{Best-level OFIs}} & \multicolumn{2}{c}{{Integrated OFIs}} & \multicolumn{2}{c}{{Returns}}\\
     \cmidrule(lr){2-3}\cmidrule(lr){4-5}\cmidrule(lr){6-7}
     
 & $\textbf{FPI}^{[1]}$  & $\textbf{FCI}^{[1]}$ & $\textbf{FPI}^{I}$ & $\textbf{FCI}^{I}$ & $\textbf{AR}$ & $\textbf{CAR}$ \\\midrule
PnL  & 0.21 & 0.43 & 0.23 & 0.39 & 0.23 & 0.40 \\
& (0.12) & (0.17) & (0.13) & (0.19) & (0.13) & (0.18) \\


\bottomrule
\end{tabular}}
\caption*{\textit{Note:} The table reports the mean values and standard deviations (in parentheses) of {annualized PnLs} of forecast-implied trading strategy of various models for forecasting one-minute-ahead returns. The predictive models include $\textbf{FPI}^{[1]}$ (Eqn \eqref{eq:for_pi}), $\textbf{FCI}^{[1]}$ (Eqn \eqref{eq:for_ci}), $\textbf{FPI}^{I}$ (Eqn \eqref{eq:for_pi_int}), $\textbf{FCI}^{I}$ (Eqn \eqref{eq:for_ci_int}), \textbf{AR} (Eqn \eqref{eq:for_ar}) and \textbf{CAR} (Eqn \eqref{eq:for_cr}). These statistics are averaged over 2017-2019.}
\label{tab:pi_ci_for_eco}
\end{table}

Table \ref{tab:pi_ci_for_eco} compares the performance (annualized PnL) of the forecast-implied strategies, based on forecast returns from various predictive models. It is worth noting that in the following analysis, the strategy {ignores trading costs}, as this is not the focus of our paper. Table \ref{tab:pi_ci_for_eco} shows that portfolios based on forecasts of the forward-looking cross-impact model outperform those based on forecasts of the forward-looking price impact model.

\subsection{Longer forecasting horizons} \label{sec_ci:forecast_longer}
One-minute-ahead return forecasts are not the only time horizon of interest to practitioners and academics. Additionally, we evaluate the performance of the above models and examine the forecasting ability of cross-impact terms over longer prediction horizons.

\begin{figure}[H]
    \centering
    \includegraphics[width=0.68\textwidth, trim=3mm 1mm 1cm 1cm,clip]{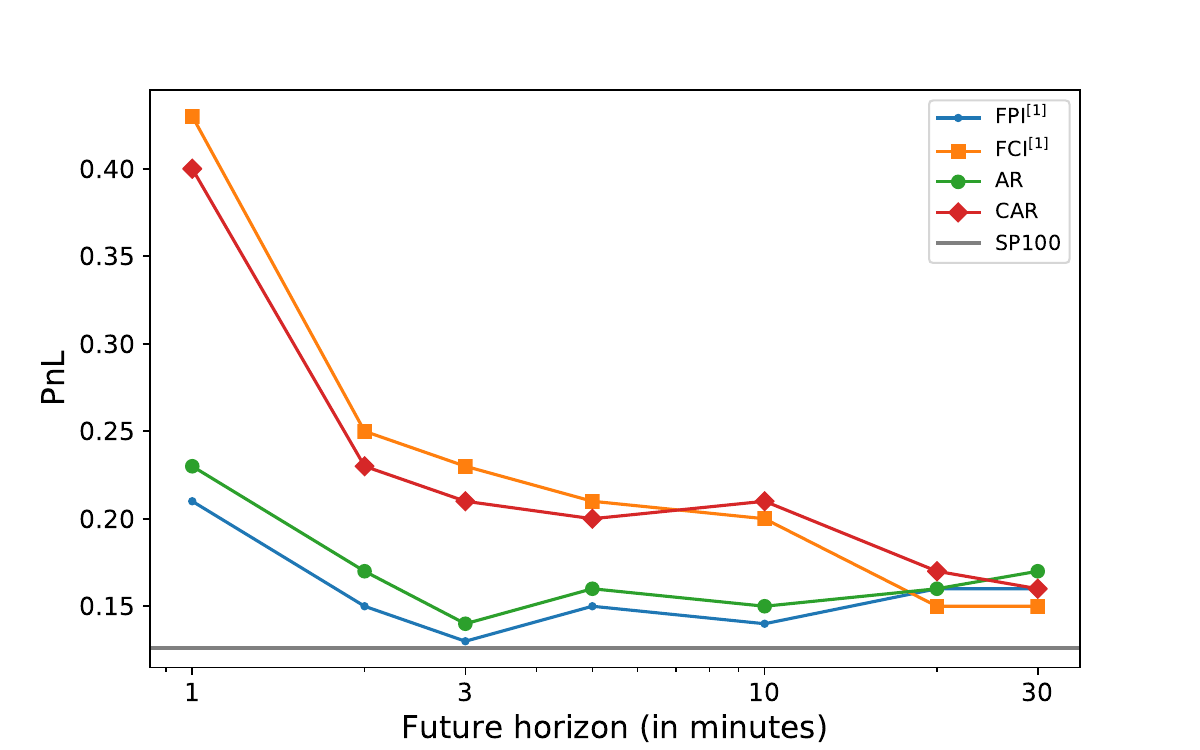}
    \caption{Annualized PnL as a function of the forecasting horizon.}
    \caption*{\textit{Note:}  The $x$-axis represents the prediction horizon (in minutes), while the $y$-axis represents the annualized PnL. The grey horizontal line is the performance of the S\&P100 ETF index.}
    \label{fig:pnl_multi_horizon}
\end{figure}

Figure \ref{fig:pnl_multi_horizon} illustrates the model predictability from the perspective of raw annualized PnL across multiple horizons.\footnote{To plot this figure, we only accumulate the PnLs between [10:31, 15:30], which is the shared trading period for the studied forecasting horizons.}   Due to the similar performance of $\textbf{FPI}^{[1]}$ and $\textbf{FPI}^{I}$ (respectively, $\textbf{FCI}^{[1]}$ and $\textbf{FCI}^{I}$) over longer horizons, we only show the curves of $\textbf{FPI}^{[1]}$, $\textbf{FCI}^{I}$, \textbf{AR}, \textbf{CAR}, and a benchmark (S\&P100 ETF). It appears that superior forecasting ability arises from cross-asset terms at short horizons. However, the PnL of cross-asset models declines more quickly over longer horizons. A further study with more focus on the reasons for the predictability of cross-asset OFIs over multiple horizons is therefore suggested. Finally, the models in which each stock only relies on its own returns/OFIs marginally outperform their counterparts which use the entire cross-sectional predictors. 

\subsection{Discussion about predictive cross-impact}

Tables \ref{tab:pi_ci_for} and \ref{tab:pi_ci_for_eco} reveal that, in contrast to the price impact model, multi-asset OFIs can provide considerably more additional explanatory power for \textit{future returns} compared to \textit{contemporaneous returns}. A possible explanation for this asymmetric phenomenon is that there exists a time lag between when the OFIs of a given stock are formed (a so-called \textit{flow formation period}) and the actual time when traders notice this change of flow and incorporate it into their trading model (see \citet{buccheri2021high}).\footnote{{A closely-related phenomenon is the Epps effect documented by \citet{epps1979comovements}, which showed that the empirical correlation estimates tend to decrease when the sampling is done at high frequencies. Previous research (including \citet{reno2003closer, toth2009epps, zhang2011estimating}) demonstrated that the Epps effect might be explained by the non-synchronicity, the possible lead-lag relationship between stock returns, etc. \citet{toth2009epps} also described that the Epps effect might be caused by the reaction time of traders to news and events, which is usually spread out over a time interval of a few minutes.}} For example, assume a trader submitted an unexpectedly large amount of buy limit orders of Apple (AAPL) at 10:00 am, at either the first level or potentially deeper in the book. Other traders may notice this anomaly and adjust their portfolios (including Apple) at a later time, for example, 10:01 am. In this case, the OFIs of Apple may indicate future price changes of other stocks. 

Consistent with our explanation, \citet{hou2007industry} argued that the gradual diffusion of industry information is a leading cause of the lead-lag effect in stock returns. \citet{cohen2008economic} found that certain stock prices do not promptly incorporate news pertaining to economically related firms, due to the presence of investors subject to attention constraints. Further research should be undertaken to investigate the origins of the success of multi-asset OFIs in predicting future returns.

It is also interesting to note that forward-looking models using integrated OFIs cannot significantly outperform models using the best-level OFIs. This phenomenon might stem from the fact that the integrated OFIs 
do not explicitly take into account the level information (distance of a given level to the best bid/ask) of multi-level OFIs, and are agnostic to different sizes resting at different levels on the bid and ask sides of the book. Previous studies (such as  \citet{hasbrouck2002limit,cao2009information, cenesizoglu2022asymmetric}) demonstrated that traders might strategically choose to place their orders in different levels of the book depending on various factors, therefore limit orders at different price levels may contain different information content with respect to predicting future returns. A further study with more focus on the impact of multi-level OFIs over different time horizons is suggested. 

\section{Conclusion}\label{sec_ci:conclusion}
We have systematically examined the impact of OFIs from multiple perspectives. The main contributions can be summarized as follows. 

First, we verify the effects of multi-level and cross-asset OFIs on contemporaneous price dynamics. We introduce a new procedure to examine the cross-impact on contemporaneous returns. Under the sparsity assumption of cross-impact coefficients, we use LASSO to describe such a structure and compare the performances with the price impact model which only utilizes a stock's own OFIs. We implement models with the best-level OFIs and integrated OFIs, respectively. The results first demonstrate that our integrated OFIs provide a higher explanatory power for price movements than the widely-used best-level OFIs. More interestingly, in comparison with the price impact model using best-level OFIs, the cross-impact model exhibits additional explanatory power. However, the cross-impact model with integrated OFIs cannot provide extra explanatory power to the price impact model with integrated OFIs, indicating the effectiveness of our integrated OFIs.

In addition, we apply the price impact and cross-impact models to the challenging task of predicting future returns. The results reveal that involving cross-asset OFIs can increase out-of-sample $R^2$. We subsequently demonstrate that this increase in out-of-sample $R^2$ leads to additional economic profits, when incorporated in common trading strategies, thus providing evidence of cross-impact over short future horizons. We also find that predictability of cross-impact terms vanishes quickly over longer horizons.

\paragraph{Future research directions.} 
There are a number of interesting avenues to explore in future research. One such direction pertains to the assessment of whether cross-asset {multi-level} OFIs can improve the forecast of future returns (in the present work, we only considered the best-level OFI and integrated OFI due to limited computing power).   
{Another interesting direction pertains to performing a similar analysis as in the present paper, but for the last 15-30 minutes of the trading day, where a significant fraction of the total daily trading volume occurs. For example, for the first few months of 2020 in the US equity market, about 23\% of trading volume in the 3,000 largest stocks by market value has taken place after 3:30 pm, compared with about 4\% from 12:30 pm to 1 pm (\citet{WSJ_last30mins}). It would be an interesting study to explore the interplay between the OFI dynamics and this surge of trading activity at the end of U.S. market hours.}


\bibliographystyle{plainnat}
\bibliography{ref}


\appendix
\renewcommand\thefigure{\thesection.\arabic{figure}}    
\renewcommand\thetable{\thesection.\arabic{table}}    
\setcounter{figure}{0}    
\setcounter{table}{0}
\section{Aggregation of multi-level OFIs} \label{sec_ci:integrate}


{Table \ref{tab:evr_ofi_multi} presents evidence of the effectiveness of PCA in selecting weights for combining multi-level OFIs. However, Figure \ref{fig:PC1_OFI_detail} shows that the weights derived from PCA are not extremely different. This prompts us to consider a simpler method, namely the simple average (SA), to achieve similar performance. Table \ref{tab:evr_ew_ofi} reveals that the explained variance of the SA increases as more levels are included. Nonetheless, the SA across 10 levels is inferior to the first principal component (PC) in terms of EVR, i.e. 85.07\% of SA vs 89.06\% of PC in Table \ref{tab:evr_ofi_multi}. Additionally, Table \ref{tab:evr_ew_detail} provides further evidence that PC consistently performs better than SA across different subsets grouped by stock-specific characteristics.}


\begin{table}[H]
    \centering
    \caption{Average percentage and the standard deviation (in parentheses) of variance attributed to SA across multiple levels.} 
    \resizebox{1.0\textwidth}{!}{\begin{tabular}{lcccccccccc}\toprule
Average across & 1     & 2   & 3  & 4  & 5  & 6  & 7  & 8  & 9  & 10  \\\midrule
Explained Variance Ratio & 13.82 & 22.81 & 32.53 & 42.73 & 52.58 & 61.45 & 68.82 & 75.10 & 80.49 & 85.07 \\
 & (9.04) & (10.09) & (10.35) & (10.17) & (9.58) & (8.69) & (7.71) & (7.10) & (7.23) & (8.14) \\\bottomrule

\end{tabular}
}
    \caption*{\textit{Notes:} The table reports the ratio (in percentage points) between the variance of SA across multiple levels and the total variance averaged across each stock and trading day.}
    \label{tab:evr_ew_ofi}
\end{table}

\begin{table}[H]
    \centering
    \caption{Explained variance ratio by different integration ways of multi-level OFIs, sorted by stock characteristics.} 
    \resizebox{0.54\textwidth}{!}{\begin{tabular}{llllll}\toprule
   &   & Volume & Volatility & Spread \\\midrule
\multirow{2}{*}{$[0\%, 25\%)$} & SA  & 83.02  & 85.09      & 81.84 \\
& PC & 85.79  & 89.64      & 89.90 \\\midrule
\multirow{2}{*}{$[25\%, 50\%)$} & SA  & 87.33  & 84.75      & 87.97 \\
 & PC & 89.93  & 89.12      & 91.17 \\\midrule
\multirow{2}{*}{$[50\%, 75\%)$}  & SA  & 86.93  & 85.23      & 87.57 \\
& PC  & 90.76  & 89.03      & 89.81  \\\midrule
\multirow{2}{*}{$[75\%, 100\%]$} & SA  & 83.55  & 85.23      & 83.03  \\
 & PC  & 90.23  & 88.48      & 85.53  \\
 \bottomrule
\end{tabular}
}
    \caption*{\textit{Notes:} SA: taking the simple average of OFIs across 10 levels. PC: taking the first principal component of OFIs across 10 levels. Volume: trading volume on the previous trading day. Volatility: volatility of one-minute returns during the previous trading day. Spread: average bid-ask spread during the previous trading day. 
    $[0\%, 25\%)$, respectively $[75\%, 100\%]$, denote the subset of stocks with the lowest, respectively highest, 25\% values for a given stock characteristic.}
    \label{tab:evr_ew_detail}
\end{table}

{We then perform the price impact and cross-impact analysis with simple average multi-level OFIs, denoted as $\textbf{PI}^{SA}$ and $\textbf{CI}^{SA}$, respectively. As shown in Table \ref{tab:pi_ew}, $\textbf{PI}^{SA}$ has a similar performance with $\textbf{CI}^{SA}$, consistent with our main analysis. This again confirms that as long as multi-level orders are taken into account, adding cross-impact terms cannot significantly improve model performance. On the other hand,  $\textbf{PI}^{I}$ is slightly better than $\textbf{PI}^{SA}$. A future research direction might be to devise various weighting schemes that average the OFI information across the multiple levels, where the weights could be given, for example, by an inverse function of the distance of each price level to the mid-price or applying tensor-SVD/PCA on this data.}

\begin{table}[H]
    \centering
    \caption{Out-of-sample performance of models based on different aggregations of multi-level OFIs.} 
    \resizebox{0.5\textwidth}{!}{\begin{tabular}{lcccc}
\toprule
 &   $\textbf{PI}^{SA}$ & $\textbf{CI}^{SA}$ &  $\textbf{PI}^{I}$  &  $\textbf{CI}^{I}$ \\\midrule
OS $R^2$ & 82.34 & 82.51 &  83.83 &  83.62 \\
& (18.02) & (14.27) & (16.90) & (14.53) \\
\bottomrule
\end{tabular}
}
    \caption*{\textit{Notes:} $\textbf{PI}^{SA}$ (resp. $\textbf{CI}^{SA}$): price impact (resp. cross-impact) model using the simple average of OFIs across 10 levels.}
    \label{tab:pi_ew}
\end{table}

\setcounter{figure}{0}    
\setcounter{table}{0}
\section{Contemporaneous price impact of multi-level OFIs} \label{sec_ci:pi_multi}

To explicitly identify the impact of deeper-level OFIs, we also consider an extended version of $\textbf{PI}^{[1]}$ by incorporating multi-level OFIs as features {in the model}
\begin{equation}\label{eq:pi_deep}
\textbf{PI}^{[m]}:\quad r_{i, t}^{(h)} = \alpha_{i}^{[m]} + \sum_{k=1}^{m} \beta_{i}^{[m], k} \mathrm{ofi}_{i, t}^{k, (h)}+\epsilon_{i, t}^{[m]}.
\end{equation} 
Recall that $\mathrm{ofi}_{i, t}^{k, (h)}$ is the OFI at level $k$. We refer to this model as $\textbf{PI}^{[m]}$, and use OLS to estimate it.

The top panel of Table \ref{tab:pi_deep} shows that the in-sample $R^2$ values increase as more multi-level OFIs are included as features, which is not surprising given that $\textbf{PI}^{[m]}$ is a nested model of $\textbf{PI}^{[m+1]}$. However the increments of the in-sample $R^2$ are descending,  indicating that  much deeper LOB data might be unable to provide additional information. This argument is confirmed by the  models' performance on out-of-sample data, as shown at the bottom panel of Table \ref{tab:pi_deep}. Out-of-sample $R^2$ reaches a peak at $\textbf{PI}^{[8]}$. 

\begin{table}[H]
    \centering
    \caption{Performance of price impact models with multi-level OFIs.}
    \resizebox{1.0\textwidth}{!}{\begin{tabular}{lcccccccccc}\toprule
&  $\textbf{PI}^{[1]}$  & $\textbf{PI}^{[2]}$  & $\textbf{PI}^{[3]}$ & $\textbf{PI}^{[4]}$  & $\textbf{PI}^{[5]}$ & $\textbf{PI}^{[6]}$  & $\textbf{PI}^{[7]}$ & $\textbf{PI}^{[8]}$  & $\textbf{PI}^{[9]}$ & $\textbf{PI}^{[10]}$ \\\midrule
{IS $R^2$} & 71.16  & 81.61   & 85.07  & 86.69  & 87.66  & 88.30  & 88.74  & 89.04  & 89.24   & {89.38} \\
& (13.80)  & (11.80)   & (10.76)  & (10.30)  & (10.05)  & (9.86)  & (9.71)  & (9.57)  & (9.45)  & (9.34) \\\midrule

{OS $R^2$} & 64.64  & 75.81  & 79.47  & 81.13  & 82.05  & 82.65  & 83.01  & {83.16}  & 83.15   & 83.11 \\
& (21.82)  & (19.83)   & (18.87)  & (18.61)  & (18.58)  & (18.65)  & (18.78)  & (18.93)  & (19.49)  & (20.93) \\
\bottomrule
\end{tabular}




}
    \caption*{\textit{Note:} The table reports the mean values and standard deviations (in parentheses) of both {in-sample} and out-of-sample $R^2$ (in percentage points) of $\textbf{PI}^{[m]}$ ($m=1, \dots, 10$) when modeling contemporaneous returns. These statistics are averaged across each stock and each regression window.}
    \label{tab:pi_deep}
\end{table}

\begin{figure}[H]
    \centering
    \subfigure[Average \label{fig:coef_deep_aver}]{
    \includegraphics[width=.23\textwidth, trim=0cm 0cm 1cm 1cm,clip]{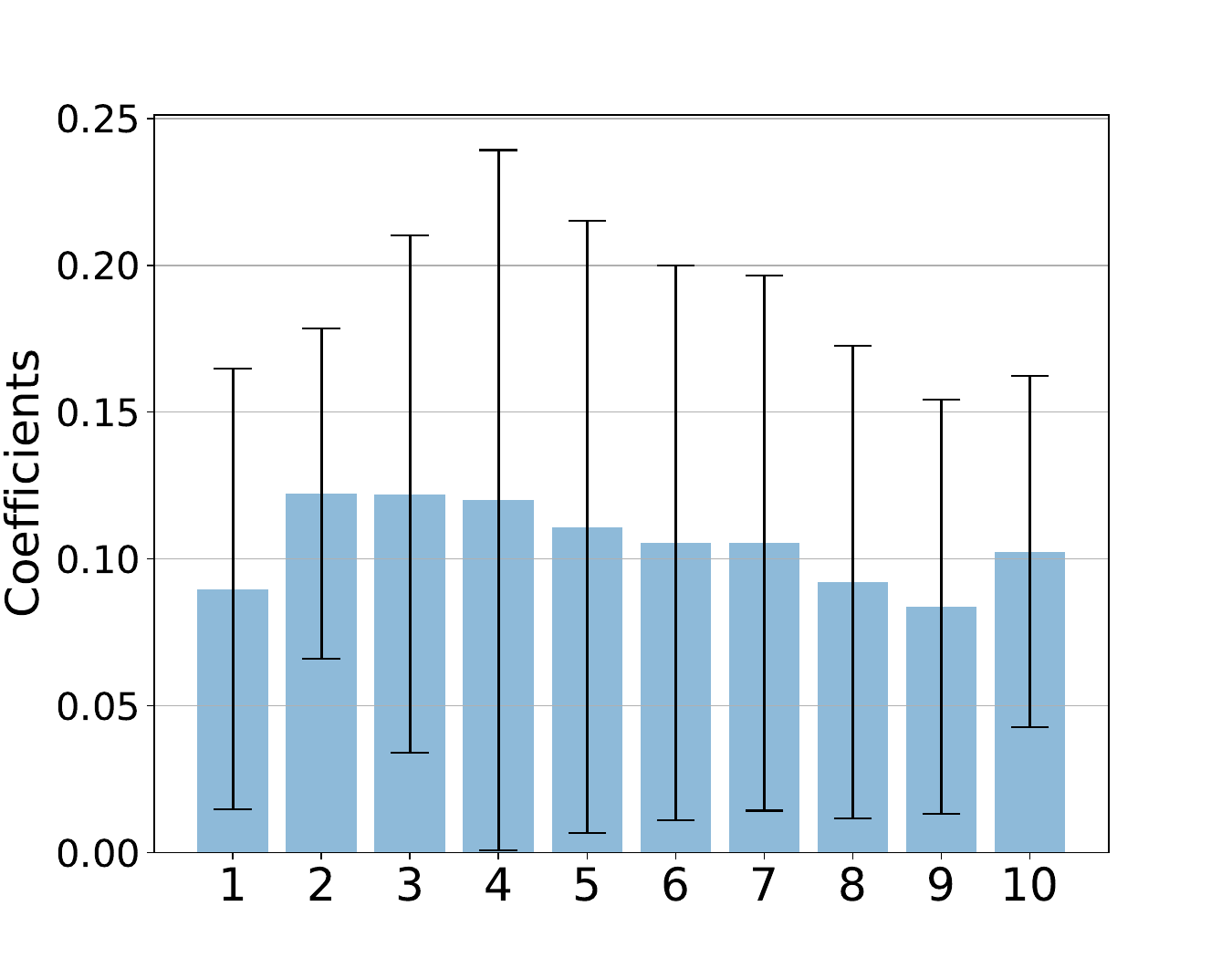}
    }
    \subfigure[Volume]{
    \includegraphics[width=.22\textwidth, trim=0cm 0cm 1cm 1cm,clip]{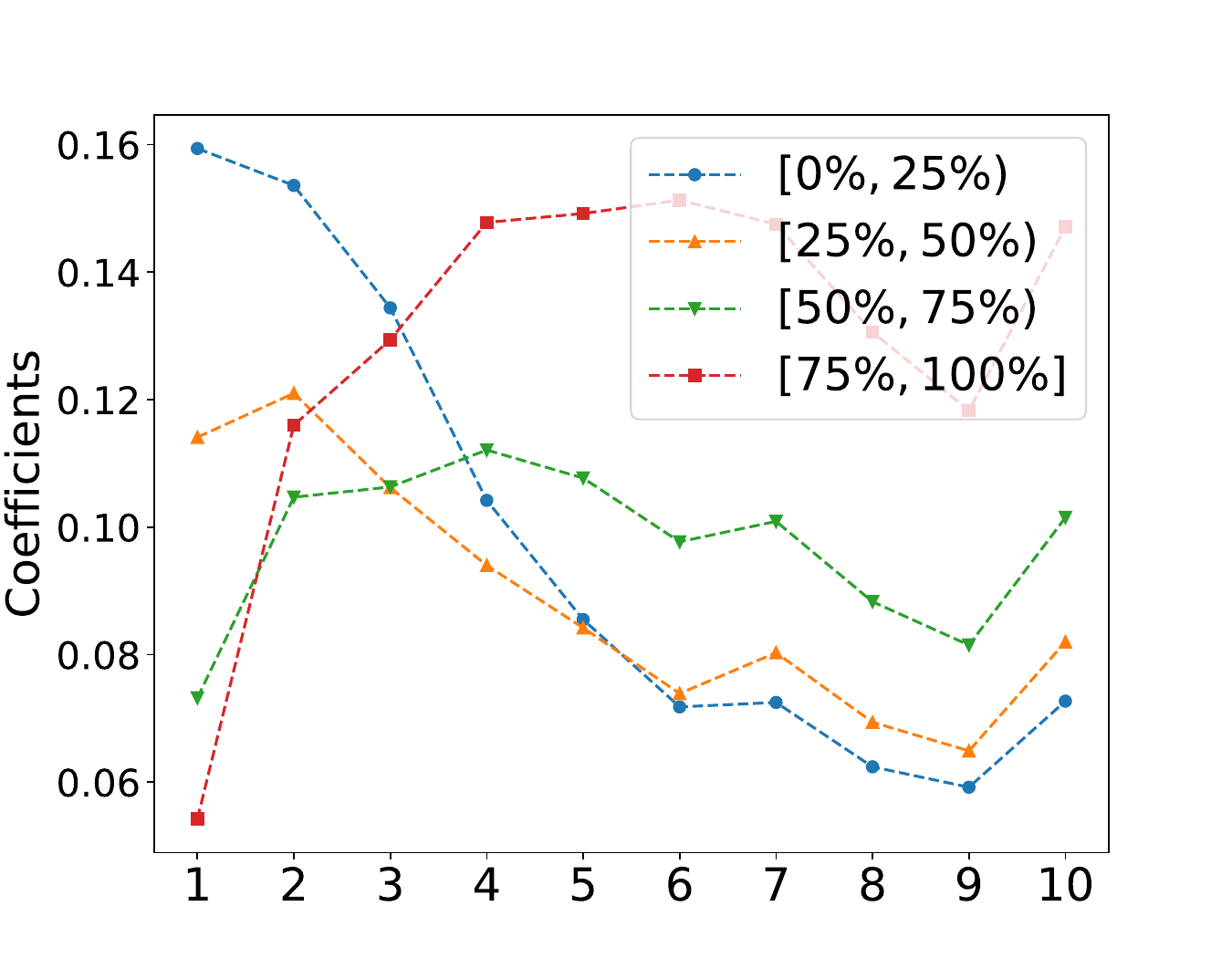}
    }
    \subfigure[Volatility]{
    \includegraphics[width=.22\textwidth, trim=0cm 0cm 1cm 1cm,clip]{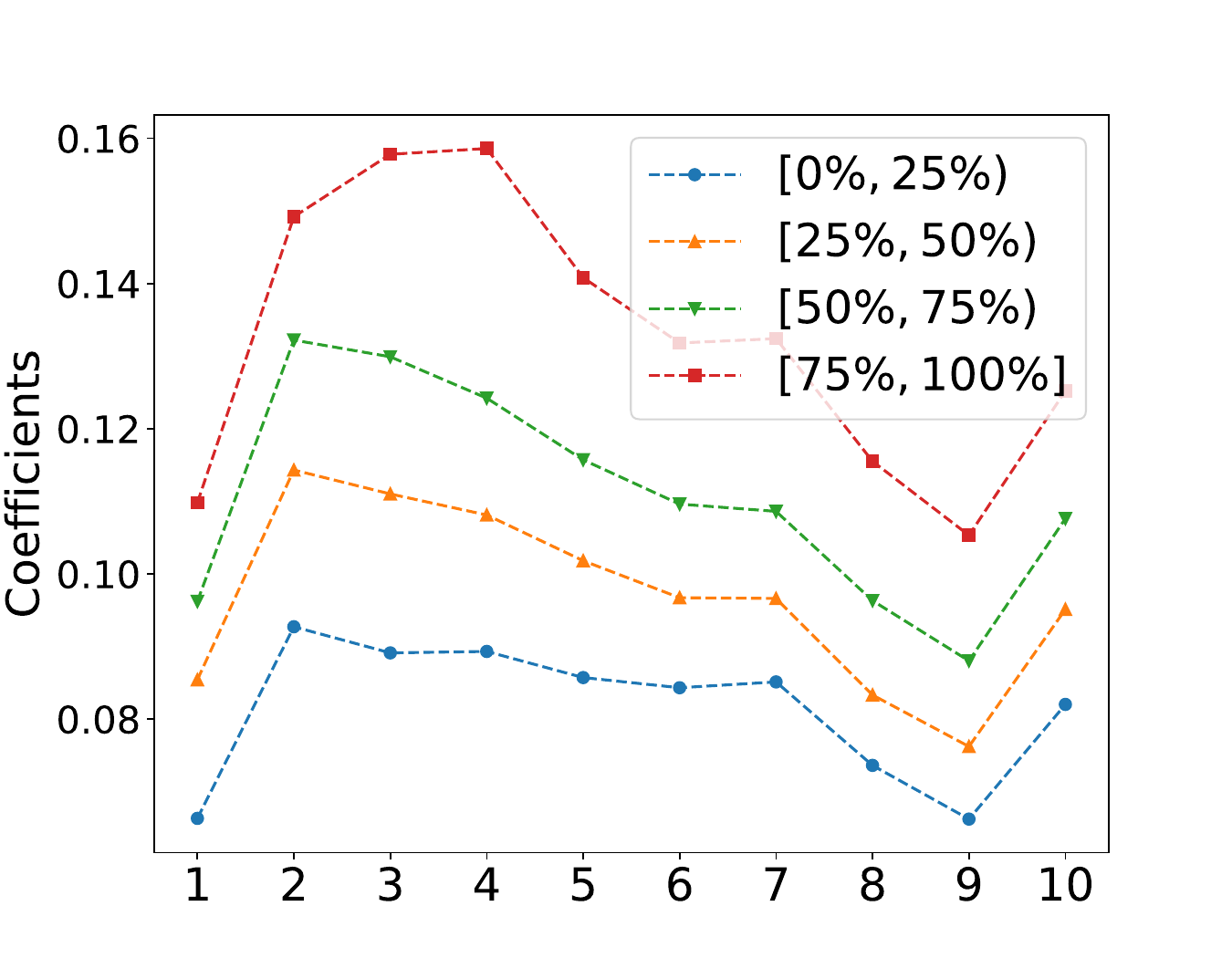}
    }
    \subfigure[Spread]{
    \includegraphics[width=.22\textwidth, trim=0cm 0cm 1cm 1cm,clip]{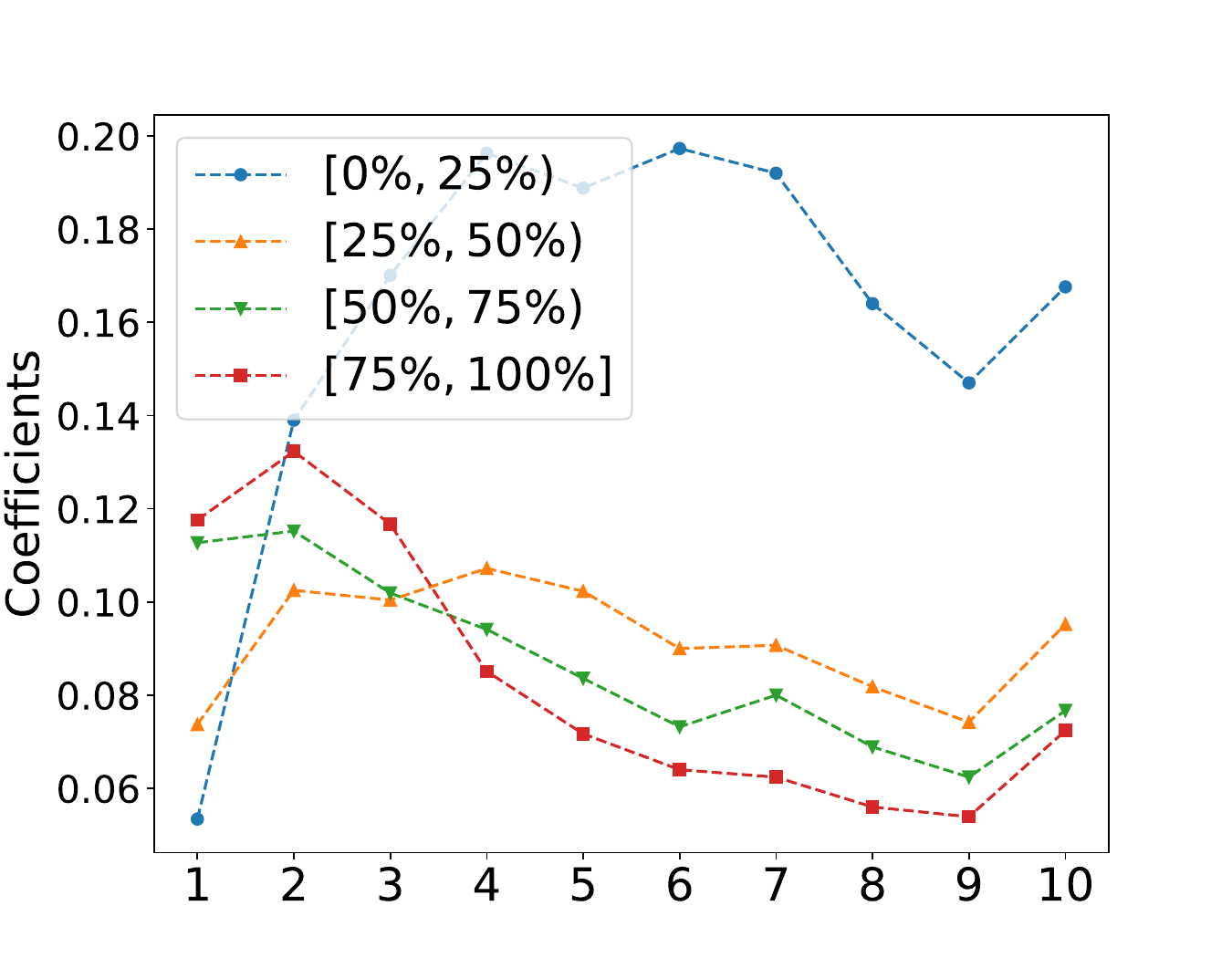}
    }
    \vspace{-2mm}
\caption{Coefficients of the model $\textbf{PI}^{[10]}$.}
\caption*{\textit{Note:} Plot (a) reports average coefficients and one standard deviation (error bars); Plots (b)-(d) show coefficients sorted by stock characteristics. Volume: trading volume on the previous trading day. Volatility: volatility of one-minute returns during the previous trading day. Spread: average bid-ask spread during the previous trading day.  $[0\%, 25\%)$, respectively $[75\%, 100\%]$, denote the subset of stocks with the lowest, respectively highest, 25\% values for a given stock characteristic. The $x$-axis represents different levels of OFIs and the $y$-axis represents the coefficients.}
\label{fig:coef_deep_detail}
\end{figure}

\paragraph{Impact comparison between multi-level OFIs.} An interesting question is whether the OFIs at different price levels contribute evenly in terms of price impact. Based on Figure \ref{fig:coef_deep_aver}, we conclude that multi-level OFIs have different contributions to price movements. Generally, OFIs at the second-best level manifest greater influence than OFIs at the best level in model $\textbf{PI}^{[10]}$, which is perhaps counter-intuitive, at first sight.

We further investigate how the coefficients vary across stocks with different characteristics, such as volume, volatility, and bid-ask spread. Figure \ref{fig:coef_deep_detail} (b)-(d) reveals that for stocks with \textit{high-volume} and \textit{small-spread}, 
order flow posted deeper in the LOB has more influence on price movements. The results regarding spread are in line with \citet{xu2018multi}, where it is observed that for large-spread stocks (AMZN, TSLA, and NFLX), the coefficients of $\text{ofi}^{m}$ (OFIs at the $m$-th level) tend to get smaller as the LOB level $m$ increases,  while for small-spread stocks (ORCL, CSCO, and MU), the coefficients of $\text{ofi}^{m}$ may become larger as $m$ increases. 

\citet{cont2014price} concluded that the effect of $\text{ofi}^{m} (m \geq 2)$ on price changes is only second-order or null. There are two likely causes for the differences between their findings and ours. First, the data used in \citet{cont2014price} includes 50 stocks (randomly picked from S\&P500 constituents) for a single month in 2010, while we use the top 100 large-cap stocks {for 36 months  during} 2017-2019. Second, \citet{cont2014price} considered the average of the coefficients across 50 stocks. In our work, we first group 100 stocks by firm characteristics, and then study the average coefficients of each subset. Therefore, our results are based on a more granular analysis, across a {significantly longer} period of time.

\setcounter{figure}{0}    
\setcounter{table}{0}
\section{Comparison with Capponi \& Cont (2020)} \label{sec_ci:ci_common}
One closely related work is \citet{capponi2020multi} (CC hereafter), where the authors propose a two-step procedure to justify the significance of cross-impact terms and render a different conclusion about cross-impact. 

In the first step, the authors use OLS to decompose each stock's OFIs ($\text{ofi}_{i, t}^{1, (h)}$) into the common factor of OFIs ($F_{\text{ofi}, t}^{(h)}$), that is the first principal component of the multi-asset order flow imbalances, and obtain the idiosyncratic components ($\tau_{i,t}^{(h)}$) of the OFIs, for each individual stock.

\vspace{-5mm}
\begin{equation}
\label{eq:common_ofi}
    \text{ofi}_{i, t}^{1, (h)} = \mu_{i} + \gamma_{i} F_{\text{ofi}, t}^{(h)}+\tau_{i, t}^{(h)}.
\end{equation}

In the second step, they regress returns ($r_{i, t}^{(h)}$) of stock $i$ against (i) the common factor of OFIs ($F_{\text{ofi}, t}^{(h)}$), (ii) the idiosyncratic components of its own OFIs ($\tau_{i,t}^{(h)}$), and (iii) the idiosyncratic components of the OFIs of other stocks ($\tau_{j, t}^{(h)}, j \neq i$). Finally, we arrive at the cross-impact model proposed by \citet{capponi2020multi} in Eqn \eqref{eq:common_ci}, denoted as $\textbf{CI}^{CC}$.

\begin{equation}
\label{eq:common_ci}
\textbf{CI}^{CC}:\quad r_{i, t}^{(h)} = \alpha_{i}^{CC} +\beta_{i0}^{CC} F_{\text{ofi}, t}^{(h)} + \sum_{j =1}^N \beta_{ij}^{CC} \tau_{j, t}^{(h)} + \eta_{i, t}^{CC}.
\end{equation}

$\textbf{CI}^{CC}$ is compared with a parsimonious model $\textbf{PI}^{CC}$ (Eqn \eqref{eq:common_pi}), in which only the common order flow factor and a stock’s own idiosyncratic OFI are utilized. 
\begin{equation}
\label{eq:common_pi}
\textbf{PI}^{CC}:\quad  r_{i, t}^{(h)} = \alpha_{i}^{CC} +\beta_{i0}^{CC} F_{\text{ofi}, t}^{(h)} + \beta_{ii}^{CC} \tau_{i, t}^{(h)}+ \epsilon_{i, t}^{CC}.
\end{equation}

We estimate the $\textbf{PI}^{CC}$ and $\textbf{CI}^{CC}$ models on historical data, under the same setting as in Section \ref{sec_ci:pi_exp}. Given that there are more features than observations, we employ LASSO in the second step to testify the intraday cross-impact of the idiosyncratic OFIs.

Similarly, we present the both in-sample and out-of-sample $R^2$ values of $\textbf{PI}^{CC}$ and $\textbf{CI}^{CC}$ in Table \ref{tab:pi_ci_common}. We observe small improvements (1.37\% in in-sample tests, 0.58\% in out-of-sample tests) from $\textbf{PI}^{CC}$ to $\textbf{CI}^{CC}$. From considering Tables \ref{tab:pi_ci_is}, \ref{tab:pi_ci_oos}, and \ref{tab:pi_ci_common}, we also observe that introducing the common factor leads to {quite small} changes in the model's explanatory power of price dynamics in the in-sample and out-of-sample tests. Moreover, the models employing integrated OFIs continually outperform others.

\begin{table}[H]
\centering
\caption{Performance of CC's models.}
\resizebox{0.33\textwidth}{!}{\begin{tabular}{lcc}
\toprule
 &   $\textbf{PI}^{CC}$ &  $\textbf{CI}^{CC}$  \\\midrule
IS $R^2$ & 72.58 &  73.95 \\
& (13.22) & (12.56) \\\midrule
OS $R^2$ & 64.78 &  65.36 \\
& (19.95) & (18.68) \\
\bottomrule
\end{tabular}

}
\caption*{\textit{Note:} The table reports the mean values and standard deviations (in parentheses) of both {in-sample} and out-of-sample $R^2$ (in percentage points) of $\textbf{PI}^{CC}$ and  $\textbf{CI}^{CC}$ when modeling contemporaneous returns. These statistics are averaged across each stock and each regression window.}
\label{tab:pi_ci_common}
\end{table}

\citet{capponi2020multi} claim that the main determinants of impact is from idiosyncratic order flow imbalance as well as a market order flow factor common across stocks;  we conclude that as long as the multi-level   OFIs are included, additional cross-impact terms are not necessary. The results also reveal that the sparse price impact model with integrated (or multi-level) OFIs can explain the price dynamics   better than the models proposed by \citet{capponi2020multi}.

\setcounter{figure}{0}    
\setcounter{table}{0}
\section{High-frequency updates of contemporaneous models}\label{sec_ci:ci_sup}
In this experiment, we use a 30-minute window to estimate contemporaneous models. We then apply the estimated coefficients to fit data in \textit{the next one minute}, and repeat this procedure every minute. The results summarized in Table \ref{tab:pi_ci_highfreq} reveal similar conclusions as in Section \ref{sec_ci:ci}, illustrating the robustness of our findings.

\begin{table}[H]
    \centering
    \caption{Performance of various models under \textit{one-minute update frequency}.}
    \resizebox{0.53\textwidth}{!}{\begin{tabular}{lcccc}\toprule
    & \multicolumn{2}{c}{{Best-level OFIs}} & \multicolumn{2}{c}{{Integrated OFIs}}\\
   \cmidrule(lr){2-3}\cmidrule(lr){4-5}
   
 & $\textbf{PI}^{[1]}$ & $\textbf{CI}^{[1]}$ & $\textbf{PI}^{I}$ & $\textbf{CI}^{I}$ \\\midrule

IS $R^2$  & 70.80 & 73.55 & 86.10  &  86.84\\
          & (13.10)   & (12.73)   & (9.64)   & (8.79) \\\midrule
OS $R^2$ & 59.67 & 61.46 & 78.88 & 78.91\\
          & (23.15)   & (18.96)   & (16.78)   & (15.02) \\
\bottomrule
\end{tabular}}
    \caption*{\textit{Note:} The table reports the mean values and standard deviations (in parentheses) of both {in-sample} and out-of-sample $R^2$ (in percentage points) of various models when modeling contemporaneous returns in {one-minute update frequency}. The models include $\textbf{PI}^{[1]}$ (Eqn \eqref{eq:pi_best}), $\textbf{CI}^{[1]}$ (Eqn \eqref{eq:ci_best}), $\textbf{PI}^{I}$ (Eqn \eqref{eq:pi_int}), and $\textbf{CI}^{I}$ (Eqn \eqref{eq:ci_int}). These statistics are averaged across each stock and each regression window.}
    \label{tab:pi_ci_highfreq}
\end{table}

\setcounter{figure}{0}    
\setcounter{table}{0}
\section{Additional results of Section \ref{sec_ci:forward}}\label{sec_ci:forward_add}




{One interesting question to consider when examining the predictability of cross-impact is whether the lead-lag effect is linked to the frequency of LOB updates. As shown in Table \ref{tab:r2_fpi_fci_detail}, the results indicate that assets with a higher frequency of book updates tend to lead ``slower'' assets, which aligns with the findings reported by \citet{kolm2021deep}.
}

\begin{table}[H]
    \centering
    \caption{Out-of-sample $R^2$ of various {predictive} models sorted by book updates.} 
    \resizebox{0.6\textwidth}{!}{\begin{tabular}{ccccc}\toprule
  &  $[0\%, 25\%)$ & $[25\%, 50\%)$  &  $[50\%, 75\%)$ & $[75\%, 100\%]$ \\\midrule
$\text{FPI}^{[1]}$ & -0.38 & -0.37 & -0.36 & -0.34 \\
$\text{FCI}^{[1]}$ & -0.12 & -0.11 & -0.10 & -0.09 \\\midrule
$\text{FPI}^{I}$   & -0.37 & -0.35 & -0.35 & -0.33 \\
$\text{FCI}^{I}$   & -0.12 & -0.11 & -0.10 & -0.09 \\
\bottomrule
\end{tabular}}
    \caption*{\textit{Notes:} $[0\%, 25\%)$, respectively $[75\%, 100\%]$, denote the subset of stocks with the lowest, respectively highest, 25\% values according to the frequencies of book updates.}
    \label{tab:r2_fpi_fci_detail}
\end{table}

\begin{figure}[H]
    \centering
\subfigure[Coefficient matrix of $\textbf{FCI}^{[1]}$]{
\includegraphics[width=1.0\textwidth, trim=10cm 10cm 3cm 6cm,clip]{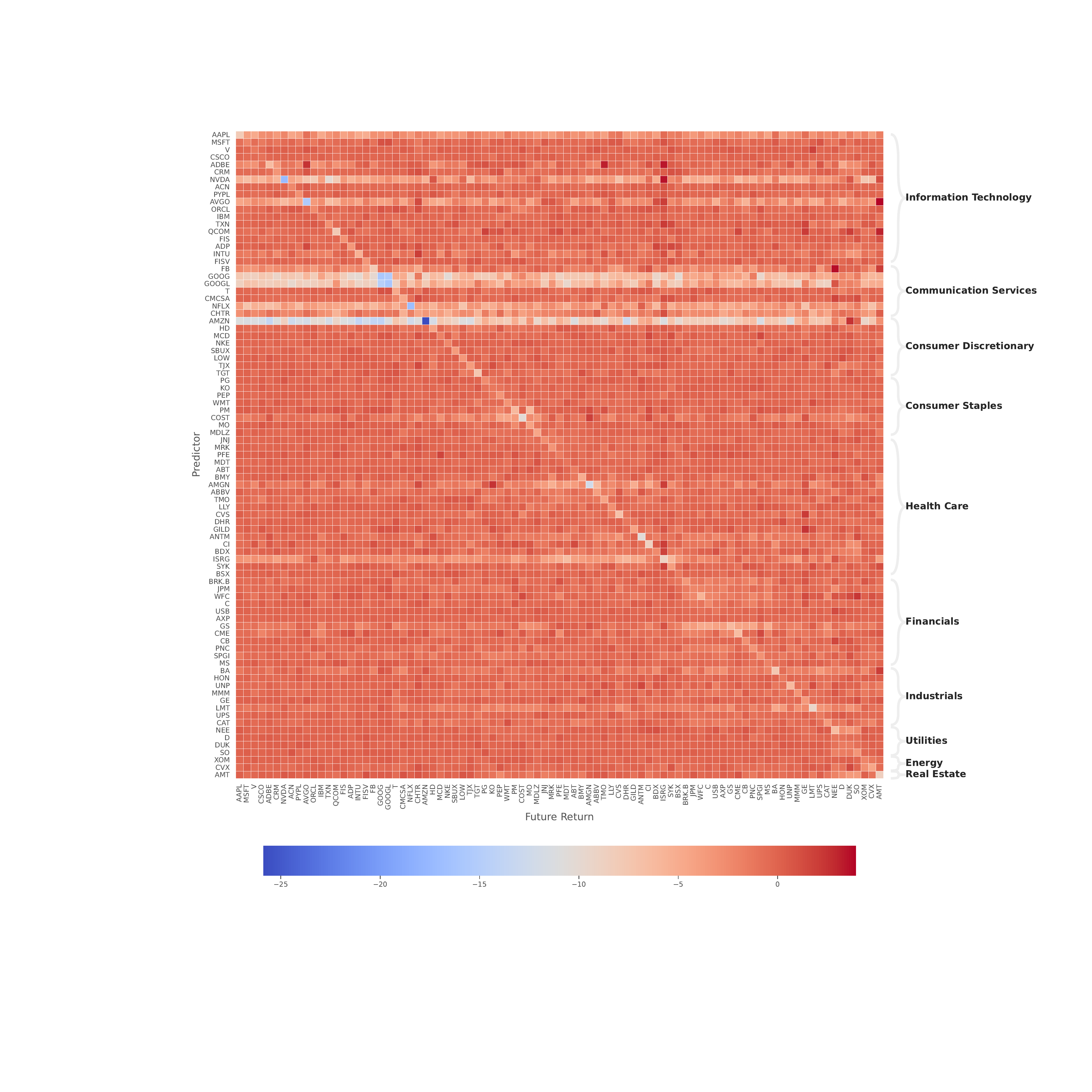}
}
\end{figure}

\begin{figure}
    \centering
\subfigure[Coefficient matrix of $\textbf{FCI}^{I}$]{
\includegraphics[width=1.0\textwidth, trim=10cm 10cm 3cm 6cm,clip]{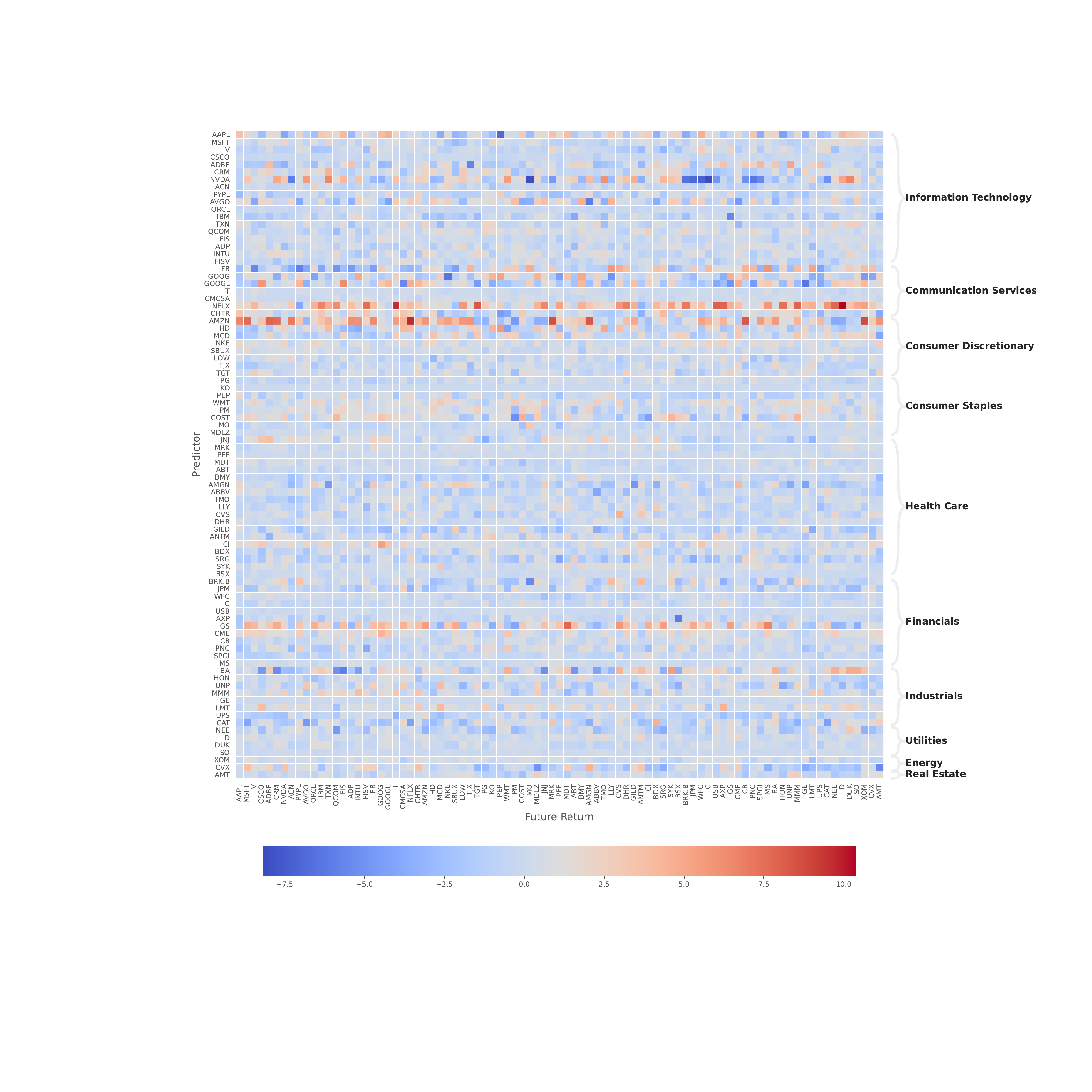}
}
\end{figure}

\begin{figure}
    \centering
\subfigure[Coefficient matrix of \textbf{CAR}]{
\includegraphics[width=1.0\textwidth, trim=10cm 10cm 3cm 6cm,clip]{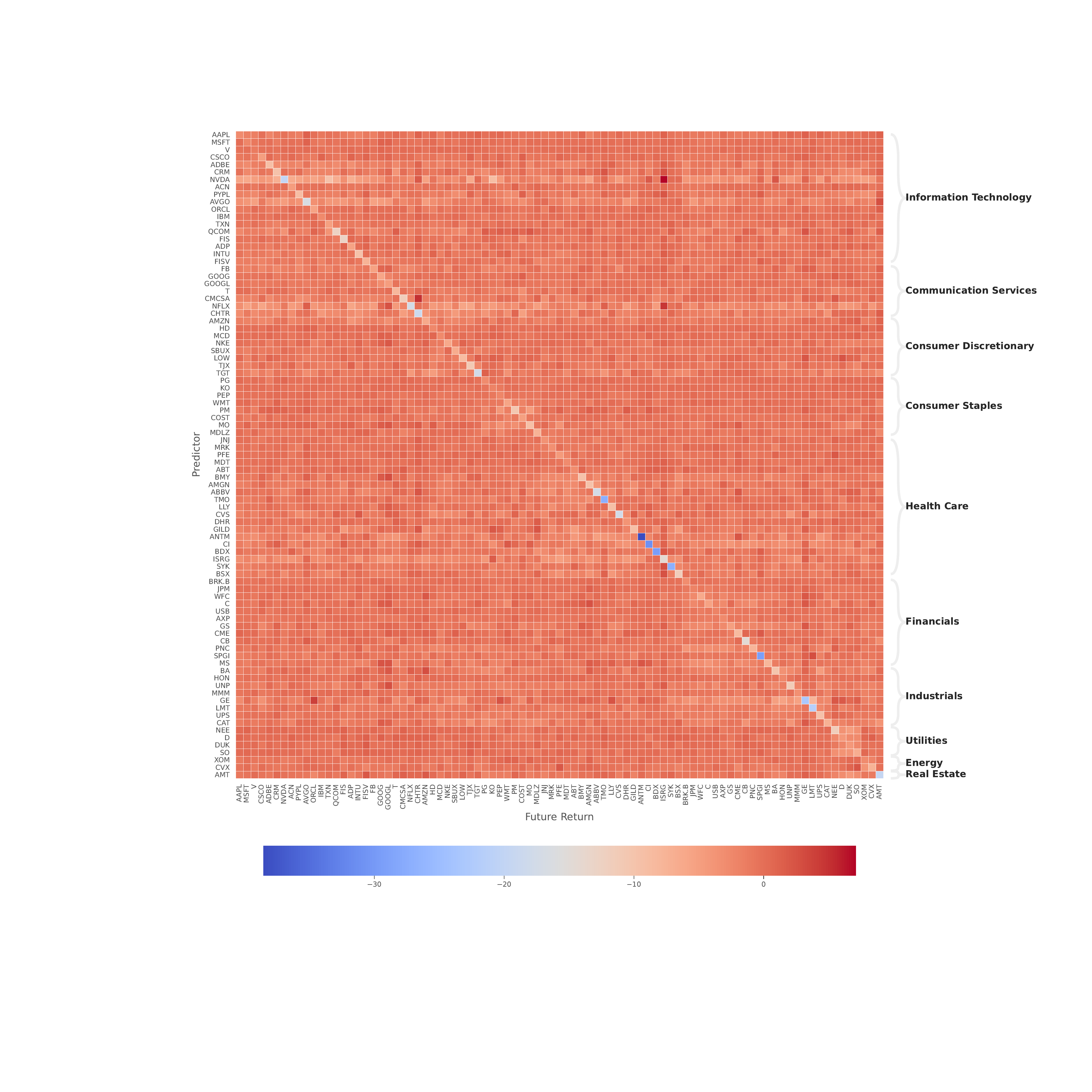}
}
\caption{Average coefficient matrices constructed from {forward-looking cross-impact} models.}
\label{fig:ci_for_coef_sec}
\end{figure}

\end{document}